\documentclass[a4paper,twoside]{amsproc}

\usepackage{snapshot}

\usepackage[a4paper,text={165mm,255mm},centering,headsep=5mm,footskip=10mm]{geometry}

\numberwithin{equation}{section}

\usepackage{graphicx}
  \graphicspath{{./figures/}}

\usepackage{natbib}

\usepackage{amssymb}

\usepackage{tensorNotation}
\usepackage{algpseudocode}
\usepackage{algorithm}
\usepackage{setspace}
\usepackage{pgfplots}

\usepackage{array} 
\usepackage{multirow} 
\usepackage{rotating} 
\usepackage{slashbox} 
\usepackage{expdlist} 
\usepackage{enumitem} 
  \def\dscbasicfont{\normalfont\sffamily}
  \renewcommand*{\descriptionlabel}[1]{\hspace\labelsep\normalfont\sffamily #1}
  \setdescription{font=\dscbasicfont}
\usepackage[olditem,oldenum]{paralist} 
\usepackage{tabularx} 
\usepackage{bigstrut}
\usepackage{longtable} 
\usepackage{threeparttable} 
  \def\TPTtagStyle#1{(#1)}
\usepackage{booktabs} 
\usepackage{amsmath}
  \DeclareMathOperator{\sgn}{sgn}

\usepackage{ragged2e}
\usepackage{caption}
\usepackage[size=footnotesize]{subcaption}
\begin{document}

\title[voFoam -- geometrical unsplit VoF algorithm on unstructered meshes]{voFoam -- A geometrical Volume of Fluid algorithm on arbitrary unstructured meshes with local dynamic adaptive mesh refinement using OpenFOAM}

\author{Tomislav Mari\'{c}}
\address{Mathematical Modeling and Analysis Group, Center of Smart Interfaces, Technische \mbox{Universit{\"a}t} Darmstadt, Germany}
\email{maric@csi.tu-darmstadt.de}
\thanks{The authors hereby gratefully acknowledge the financial support by the German Research Foundation (DFG) and the German Council of Science and Humanities, in the framework of the DFG Cluster of Excellence EXC 259.}

\author{Holger Marschall}
\address{Mathematical Modeling and Analysis Group, Center of Smart Interfaces, Technische \mbox{Universit{\"a}t} Darmstadt, Germany}
\email{marschall@csi.tu-darmstadt.de}

\author{Dieter Bothe}
\address{Mathematical Modeling and Analysis Group, Center of Smart Interfaces, Technische \mbox{Universit{\"a}t} Darmstadt, Germany}
\email{bothe@csi.tu-darmstadt.de}

\subjclass{flu-dyn}
\date{May 14, 2013.}



\begin{abstract}
A new parallelized unsplit geometrical Volume of Fluid (VoF) algorithm with support for arbitrary unstructured meshes and dynamic local Adaptive Mesh Refinement (AMR), as well as for two and three dimensional computation is developed. The geometrical VoF algorithm supports arbitrary unstructured meshes in order to enable computations involving flow domains of arbitrary geometrical complexity. The implementation of the method is done within the framework of the OpenFOAM library for Computational Continuum Mechanics (CCM) using the C++ programming language with modern policy based design for high program code modularity. The development of the geometrical VoF algorithm significantly extends the method base of the OpenFOAM library by geometrical volumetric flux computation for two-phase flow simulations.\\[0.5em]
For the volume fraction advection, a novel unsplit geometrical algorithm is developed, which inherently sustains volume conservation utilizing unique Lagrangian discrete trajectories located in the mesh points. This practice completely eliminates the possibility of an overlap between the flux polyhedra and hence significantly increases volume conservation. A new efficient (quadratic convergent) and accurate iterative flux correction algorithm is developed, which avoids topological changes of the flux polyhedra. %
Our geometrical VoF algorithm is dimension agnostic, providing automatic support for both 2D and 3D computations, following the established practice in OpenFOAM. %
The geometrical algorithm used for the volume fraction transport has been extended to support dynamic local AMR available in OpenFOAM. Furthermore, the existing dynamic mesh capability of OpenFOAM has been modified to support the geometrical mapping algorithm executed as a part of the dynamic local AMR cycle. The method implementation is fully parallelized using the domain decomposition approach.\\[0.5em]
The majority of the standard established test cases for verification and validation of VoF algorithms has been thoroughly tested with varying Courant numbers. Our results for the first time show a VoF algorithm on unstructured meshes to be robust, mass conservative and boundedness-preserving for complex time-varying velocity fields. The obtained volume of symmetric difference interface reconstruction errors are the lowest reported so far in the literature for unstructured meshes.
\end{abstract}

\maketitle

\section{Introduction}
\label{section:introduction}
The \emph{geometrical} Volume-of-Fluid (VoF) method is a method where the 
interface is represented by a set of simple geometrical entities and advection 
of the volume fraction is accomplished using geometrical algorithms. When
the interface is represented as a set of polygons (planes) reconstructed  from
the volume fraction field, the geometrical VoF algorithm is commonly known as
the Piecewise Linear Interface Calculation (PLIC) algorithm \citep{deBarPLIC,
RiderReconstructing1998}. Geometrical operations involving interface planes and
mesh cells are then performed in order to advect the volume fraction field.

The widespread use of the geometrical Volume of Fluid (VoF) method for
simulating two-phase flow is mostly based on its intrinsic mass conservation 
and the ability to automatically deal with topological changes of the fluid 
interface (no special operations on the discrete interface geometry are required). 
However, the majority of the developments related to the geometrical VoF method 
have so far been done on Cartesian and structured meshes because of the high 
numerical accuracy they provide and the simplicity of the algorithms relying 
on structured mesh points. Problems involving complex geometrical shapes of the 
flow domain, usually rely on automatic (fast) algorithms for generation of 
unstructured meshes. For this reason, the use of unstructured meshes has prevailed 
in those parts of the industry and research where complex geometry of the flow 
domain is of importance. However, the use of an accurate geometrical VoF method 
for such problems was not possible so far; instead VoF methods involving algebraic 
advection of the volume fraction field are in widespread use. 

This contribution utilises the OpenFOAM{\textsuperscript{\textregistered}}
software for Computational Continuum Mechanics (CCM) as an implementation
platform for the geometrical VoF method on arbitrary unstructured meshes. As 
detailed in the following, the developed geometrical VoF method is based on novel 
a directionally unsplit advection algorithm using a full discrete Lagrangian flow 
map on unstructured meshes and thus allowing for accurate direct numerical 
simulation (DNS) of two-phase flow in domains of arbitrary geometrical complexity. 
The geometrical VoF algorithm for volume fraction transport is shown to be robust, 
volume conservative and boundedness preserving; it is fully parallelized using 
the domain decomposition approach and supports dynamic local Adaptive Mesh 
Refinement (AMR). The algorithm utilizes a new efficient (quadratic convergent) and 
accurate iterative flux correction algorithm, that avoids topological changes of 
the flux polyhedra and inherently ensures mass conservation. To the best of the 
authors' knowledge, for the first time such a geometrical VoF method is shown to 
be capable to deal with three dimensional, spatially complex and time varying 
velocity fields. The obtained volume of symmetric difference interface 
reconstruction errors are the lowest reported so far in the literature for 
unstructured meshes. %
The development of the geometrical VoF algorithm significantly extends the method 
base of the OpenFOAM library using a modern policy based design for high program 
code modularity and combinatorial complexity of sub-algorithms for geometrical 
reconstruction and advection. %

OpenFOAM is a Finite Volume Method (FVM) library for Computational Continuum 
Mechanics (CCM) written in the C++ programming language. It has a modular structure 
and provides a possibility of an extensive re-use of the available functionality. 
A broad choice of mathematical and constitutive models is available, as well as 
linear solvers, dynamic mesh operations involving both mesh motion and topological
operations and a comprehensive set of boundary conditions, etc. Details of the 
development model of OpenFOAM and its relationship to continuum mechanical modeling 
and simulations of CCM problems are provided by \citet{WellerJasak1998} and
\citet{Jasak2007}. The benefit of developing the geometrical VoF method for 
unstructured  meshes using OpenFOAM within an \emph{Open Source} framework is thus 
two-fold: it provides an implementation with the ability to deal with arbitrarily 
complex flow domains, and it allows for extensions of the method by coupling with 
the present functionality of the OpenFOAM library. 

The goal of this work is to develop a geometrical VoF library and a set of
solver applications that can deal with two-phase flows involving complex
transport processes (\emph{complex flows}) in complex domains (\emph{complex
geometry}). The topology of an unstructured meshes dictates our choice of 
algorithms for both the reconstruction of the interface geometry and the advection 
of the volume fraction, as well as the underlying geometrical operations. The fluid
interface is approximated with a piecewise planar geometry which is
reconstructed using the method of \citet{ParkerYoungsGradient} with an
alternative accurate parallelized gradient calculation on arbitrary
unstructured mesh and is able to deal with arbitrary shapes of the mesh cells.
An Eulerian unsplit geometrical algorithm is chosen for the volume fraction
advection, based on a full unique Lagrangian trajectories located in the cell
corner points. We do not consider complicated geometrical operations necessary
for non-convex congruent polyhedra, which may be constructed by the advection
algorithm in those parts of the flow domain where insufficient resolution is
provided. The reason for this decision is that the flow will
be sufficiently resolved by local dynamic AMR near the interface. 
 
The accuracy of the geometrical VoF method can be enhanced
when coupled with dynamic local Adaptive Mesh Refinement (AMR)\footnote{From
this point on in the text, dynamic local AMR will be refered to as AMR since it
must follow an evolving fluid interface when applied to two-phase flow
simulations.} (see \citet{KotheUnstructured1996} and \citet{Cerne2002} for
details). Extensions to higher order geometrical methods such as the Moment-of-Fluid 
(MoF) method of \citet{AhnShashkov2009} or higher order reconstruction
(e.g.  Patterned Interface Reconstruction by \citet{MossoPIR2009}) are also
possible if it proven to be necessary (even with applied AMR).

\subsection{State of the art}

For the sake of clarity, we provide a brief overview of the important aspects
of the geometrical VoF method as well as the related terminology. Emphasis is
put upon geometrical VoF methods and related algorithms rather than the algebraic
VoF method. The interested reader is referred to appropriate references in 
literature once the scope of this overview is left. 

In order to advect the volume fraction field using geometrical algorithms, the
geometrical representation of the interface needs to be sufficiently accurate,
and the geometrical advection algorithm must conserve mass as well as avoid
artificial deformation of the interface, while keeping the volume fraction
field bounded between values of $0$ and $1$. 

Since the interface is described as a set of planes (polygons) the orientation 
of which is given by the volume fraction field, the planes are disconnected
on cell faces. Discontinuities of the interface planes at cell faces may be 
described in terms of \emph{geometrical stability} of the interface.
The discontinuities decrease with the increase of mesh
resolution and the accuracy of the gradient calculation, used to determine the
orientation of the plane normals. Furthermore, the orientations of the
interface planes may be additionally optimized in a way that does not impose
errors in mass conservation. Such second order convergent reconstruction
algorithms\footnote{Order of convergence of geometrical algorithms is related
to convergence of standard geometrical reconstruction or advection errors.}
result in smaller discontinuities, but involve additional computational
complexity. However, it is of utmost importance to reduce them to a minimum, 
since geometrical instabilities generated by the reconstruction algorithm cause
accumulation of other errors during the advection of the volume fraction field
\citep{Cerne2002}. 

Geometrical advection algorithms may be sorted into two general categories
based on the direction splitting: directional (operator) split and un-split
algorithms. Another division may be defined based on the choice of the mesh
geometry used to advect the volume fraction field: \emph{face swept}, and
\emph{Lagrangian re-mapping} algorithms. 

\emph{Operator split algorithms} compute the volume fraction fluxes
interchangeably in directions orthogonal to the coordinate axes. Such ordering
is native to a structured mesh, but can be extracted from the unstructured
hexahedral mesh as well. Since the advection directions are interchanged, each
advection iteration is followed by a reconstruction step where the interface
geometry is computed from the updated volume fraction field. 

\emph{Directional un-split algorithms} compute the advection of the volume
fraction field in a single computational step, using the velocity vectors.
Hence, a single reconstruction of the interface is required after each
advection step.

\emph{Face-swept algorithms} rely on flux calculations coming from sweeping or
\emph{back-tracing} cell faces along the streamlines given by the velocity
field. A \emph{swept polyhedron} is then adjusted, such that its volume equals
the value of the volumetric flux coming from the numerical solution of the
pressure equation in order to assure mass conservation. 

\emph{Eulerian backtracking / Lagrangian re-mapping} algorithms compute the
increase of the volume fraction on a per-cell basis, by sweeping the mesh
backward using the velocity field interpolated to the mesh vertices, and
intersecting the swept mesh with the geometrical data of the interface. After
this Lagrangian back-tracking step, a re-mapping is executed on the underlying
Eulerian mesh to set the new volume fraction values.

On unstructured meshes, both face-swept and Lagrangian-Eulerian re-mapping
algorithms are directionally un-split. Depending on the complexity of the flow
field, both face-swept and Lagrangian re-mapping algorithms encounter problems in
geometrical operations because the created polyhedra have non-planar faces and
may even be non-convex.  

A detailed review of algorithms involving two-phase flow methods on structured
meshes with emphasis on the geometrical VoF method can be found in a book on
Direct Numerical Simulations (DNS) of two-phase flows by
\citet*{Tryggvason2011direct} and in journal contributions by
\citet{RiderReconstructing1998}, \citet{ScardovelliReview1999},
\citet{ScardovelliZaleskiEILE2003}, and \citet{Aulisa2007}. For latest 
developments on Cartesian structured meshes the interested reader is 
referred to \citep{WeymouthYue2010,MencingerZunAdaptive2011,LeChenadec2013} and 
the references therein. %
A detailed numerical analysis of the errors native to the two-dimensional 
geometrical VoF method on structured meshes is described in \citet{Cerne2002}. 
So far, main contributions to geometrical algorithms for VoF methods were 
concentrated on the following aspects: improving interface normal orientation, 
improving overall mass conservation, reducing under-and-overshoots of the volume 
fraction, and removing wisps and artificial interface separation. 

In the remainder we present a brief overview of the research efforts involving 
three-dimensional calculations and geometrical VoF algorithms which support 
\emph{unstructured} domain discretization, since they are more relevant to our work. 

\citet{Swarz1996} and \citet{KotheUnstructured1996} have produced first (to our
knowledge) publications referring to a geometrical VoF method with support for
unstructured meshes. 

\citet{KotheUnstructured1996} emphasize that the motivation behind research
towards unstructured meshes is based on the requirement for arbitrary
geometrical complexity of the flow domain. Important aspects of the gradient
reconstruction for the volume fraction field is presented, resulting in a
conclusion that a wide cell stencil ($27$ cells for a hexahedral mesh) is
necessary in order to obtain geometrical stability of the reconstructed
interface. A valuable note is provided regarding the visualization of results,
which emphasizes the importance of viewing the actual polygonal geometry of the
reconstructed interface. Standard visualization practice utilises an
iso-surface representation of the interface, where the iso-value of the volume
fraction is prescribed (e.g. to a value of $0.5$). However, the use of
iso-contours for visualizing interfaces introduces artificial connectivity of
the interface e.g. for thin filaments. Besides this, the iso-contour hides all
the errors which are shown when the polygonal geometry is used to visualize the
interface: \emph{wisps} -- small differences in the volume fraction in
the bulk of the phases from values $0$ or $1$ that cause reconstructed
interface geometry to appear (cp.\ \citet{HarvieFletcherStream2000} for
details) -- or even larger artificially separated parts of the interface 
may not be shown.

\citet{Swarz1996} developed a second order convergent interface reconstruction
method for unstructured meshes. The reconstruction is based on optimizing the
orientation of the plane normals taking into account orientation of the planes
in the neighbouring interface cells. For the advection of the volume fraction,
a full Lagrangian back tracing of the finite volume mesh is performed, followed
by a geometrical intersection with the PLIC geometry, and remapping of the
volume fraction field onto the Eulerian background mesh. This advection
algorithm is directionally un-split, it relies completely on the discrete flow
map and requires a single reconstruction step per advection iteration. 

\citet{KotheVoFchar1999} provides an overview of the geometrical VoF method
developed for simulating metal casting processes in complex molds. Although a
casting process is a specialized application of the geometrical VoF method, the
problems involving the complex domain geometry and physical characteristics of
the model (e.g. significant differences in values of physical properties
accross the interface), make the requirements applicable to a general purpose
geometrical VoF method. They show preliminary results involving reconstruction
of an interface on an unstructured tetrahedral mesh.

\citet{AhnShashkovGeometricAlg2008} present geometrical algorithms used for the
reconstruction and advection step operations, with support for polyhedral
cells. They report successful three-dimensional reconstructions of complex
geometrical shapes using the method of \citet{ParkerYoungsGradient} and the
Least-squares-Volume-of-fluid-Interface-Reconstruction-Algorithm (LVIRA)
developed by \citet{PuckettLvira1991}, both modified in order to support
polyhedral cell shapes. \citet{ShashkovDyadechkoMoF2005} present three
dimensional results of the Moment of Fluid (MoF) reconstruction method which
optimizes the difference in position of the barycentric centre of the truncated
polyhedron filled with a specific phase. The MoF method can be viewed as an
extension of the geometrical VoF method by introducing additional information
in the form of barycentric centres of the truncated cells, which results in a
more accurate reconstruction algorithm (see
\citet{AhnShashkovMultiMaterialRec2007} for details on multi-material
reconstruction on polyhedral meshes). 

\citet{jofreUnstructuredPLIC} show three-dimensional results for both
geometrical reconstruction and the volume fraction advection on unstructured
finite volume mesh. The authors present the volume fraction evolution for
rotational flow using a full discrete Lagrangian flow map. Obtained errors of
the advection algorithm are close in magnitude to those reported by other
authors on structured meshes (see \citet{LiovicCVTNA2006} and
\citet{HernandezLopezPart1Advection2008}). 

\citet{MossoPIR2009} developed the geometrical reconstruction algorithm called
Patterned Interface Reconstruction (PIR) algorithm. The algorithm is of second
order of convergence (reconctructs planes exactly) on tetrahedral (and
triangular) meshes. The interface geometry is initially reconstructed using the
Youngs PLIC reconstruction algoritm followed by linear or spherical smoothing
operations based on the geometrical pattern formed by a group of interface
polygons. Apart from reconstruction algorithm of the MoF method, this is the
only reconstruction algorithm of second order that directly applies to 
arbitrary unstructured meshes. 

\citet{LiovicCVTNA2006} have developed a three-dimensional extension of the
original algorithm of Rider and Kothe on Cartesian meshes with a new accurate
reconstruction algorithm and an elegant flux divergence correction method used
for improving the mass conservation. Errors in mass conservation appear because
of overlaps between the flux polyhedra on common mesh edges. Although the
algorithm shows good results, it relies on a directed connectivity between
cells (e.g. \emph{L-shaped stencils}) that requires the use of a structured
mesh. Additional computation of directed cell connectivity would be
possible for unstructured \emph{hexahedral} meshes, but not on arbtirary
unstructured meshes.

\citet{LopezEMFPA2004} show results with a full discrete flow map as a source
for the back tracing of the flux polyhedra on a two-dimensional Cartesian mesh.
Local geometrical corrections of the flux polyhedron are applied in order to
ensure mass conservation. For the reconstruction algorithm, a compact spline
reconstruction algorithm, with splines based on the centre points of the
reconstructed interface line segments is used. This work is extended to three
dimensions, again on a Cartesian mesh, by
\citet{HernandezLopezPart1Advection2008} and \citet{LopezPart2reconstruction}.
However, the construction of swept polyhedra is based on edge centres of the
Cartesian mesh, with velocities interpolated from the staggered velocity field
mapped to the face centres. Although a notable improvement is achieved, there
is still a possibility of an overlap between the flux polyhedra around face
corners. However, the mass conservation is strictly met by an analytical
approach to adjusting the swept polyhedron volume. 

\citet{IveyMoin2012} have implemented the so-called edge-matched flux polyhedra 
algorithm in 3D (EMFPA 3D), suggested as an improvement of the unsplit VoF 
algorithm developed by \citet{HernandezLopezPart1Advection2008}. %
The algorithm locally approximates a stream tube emanating from the flux face. 
The flux face is triangulated to construct the flux polyhedron, which then 
is modified to correct the non-planarity of the faces. Divergence correction, as 
recommended by \citet{HernandezLopezPart1Advection2008}, relies on an analytical 
expression for computing the volumes of convex polyhedrons. Adjusting the flux 
polyhedron introduces overlapping of the flux polyhedra that needs to be further 
corrected to prevent the volume fraction from over-/undershooting. Unfortunately, 
a validation of the results showing quantitative information on the standard 
advection errors are not provided.



Summarizing, all of the aforementioned algorithms rely on \emph{divergence corrections}  
in some form to account for the fact that the discrete velocity field used for the
construction of the flux polyhedron is not divergence free. Besides this issue,
wisps (see \citet{HarvieFletcherStream2000} for details) are present in the
volume fraction field, and are subsequently redistributed. 

As for AMR, \citet{PopinetTreeAMR2003} developed a very advanced geometrical VoF method 
that supports both 2D and 3D calculations with AMR on structured mesh. The
domain discretization is based on an \emph{octree} data structure which enables
fast and flexible mesh refinement operations and locally accurate
differentiation.  Recently, \citet{PopinetBalancedForce2009} has shown that the
AMR plays a significant role in reducing \emph{parasitic currents}. Details on
the capabilities and applications of the octree based geometrical VoF method
are provided by \citet{FusterPopinet2009} and \citet{AgbaglahPopinet2011}. 

Fully unstructured AMR available in OpenFOAM delegates the refinement to each
cell, and is hence able to deal in a straightforward way with arbitrary cell
shapes and solution domains of arbitrary geometrical complexity. With fully
unstructured AMR, a new mesh is created with refined and unrefined regions each
time a topological operation is executed. The octree based AMR implies a flow
domain in the shape of a regular hexahedron. %

We have chosen the fully unstructured AMR since it does not restrict us to a 
specific cell shape (allows automatic mesh generation of complex solution 
domains) and it is already a part of the OpenFOAM library, which means it can 
be coupled to other dynamic mesh functionality (mesh motion, mesh deformation, 
sliding interfaces, moving reference frames, etc.). 

In the following section the governing equations of the flow are described.
Geometrical operations required by the unstructured meshes and their use in the
reconstruction and advection step of the method are presented in the third
section. In the fourth section the coupling between the numerical solution of
the governing equations and the geometrical evolution of the volume fraction is
set out. Results and discussions are presented in the two final sections.

\section{Geometrical evolution of the volume fraction field}
\label{section:geometrical-algorithm}
\label{section:geometry} 
\begin{figure}[h]
	\centering
    \def\svgwidth{0.4\columnwidth}
       {\footnotesize
        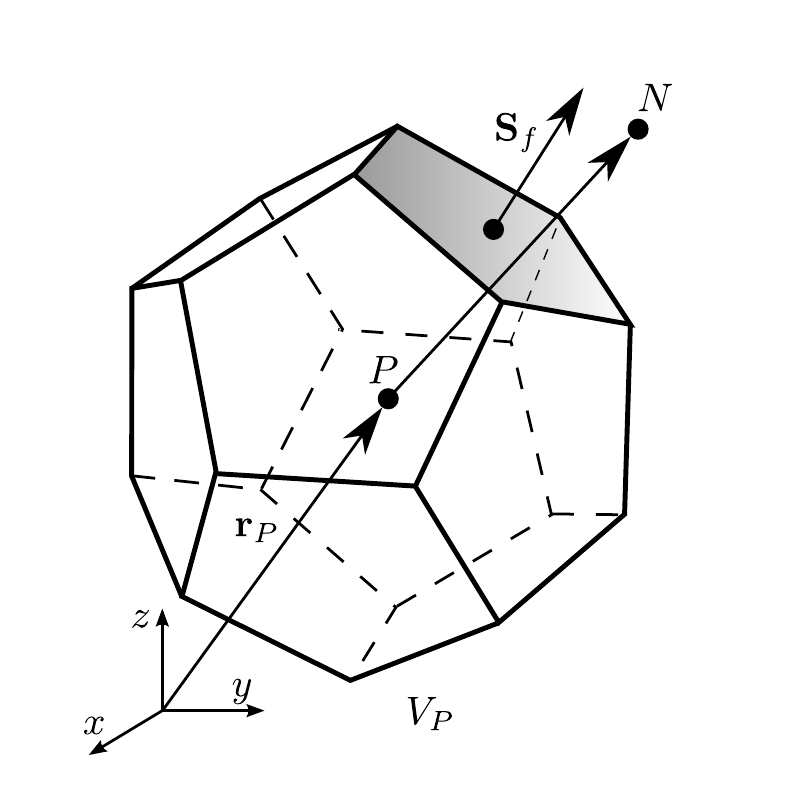
       }
	\caption{Polyhedral cell}
	\label{fig:polyhedralCell-FVM}
\end{figure}

The advection equation for the volume fraction field
of a phase $\alpha$ for an incompressible flow is written in the following
form:
\begin{equation}
  \partial_t{\alpha} + \nabla \cdot {(\U \alpha)} = 0. 
  \label{equation:alpha-advection}
\end{equation}
Applying the discretization practice of the unstructured FVM, with the first
order Euler explicit time discretization on Equation
\ref{equation:alpha-advection}, results with the following algebraic equation: 
\begin{equation}
  \alpha_{c}^n = \alpha_{c}^o - \frac{\Delta t}{V_c} 
  \sum_f  \alpha_{f} \vec{U}_f \cdot \vec{S}_f, 
  \label{equation:alpha-discrete}
\end{equation}
where $f$ denotes the face of a computational cell (cp. Fig.\ 
\ref{fig:polyhedralCell-FVM}) and, correspondingly, $\vec{U}_f$ and 
$\vec{S}_f$ are the face-centered velocity and face area normal 
vectors shown in Figure \ref{fig:polyhedralCell-FVM}. 

The unstructured FVM utilises a computational domain decomposed into
non-overlapping convex polyhedra (Figure \ref{fig:polyhedralCell-FVM}) which are the 
building elements of the unstructured mesh. However, the implementation of the mesh 
is done in an efficient way, when it comes to addressing the information needed by 
numerical algorithms. More precisely, the mesh points are defined only once as a 
set of position vectors, and are indirectly addressed by the unstructured FVM 
algorithms. The indirect addressing leads to the definitions of faces and cells 
as sets of indirect indices: the face is thus an ordered set of indices which 
relate (map) to the list of mesh points, and the mesh cell is a set of indices 
which relate to the list of mesh faces. 

The ordering of the face indices determines the orientation of the face area
normal vector $\vec{S}_f$, shown in Figure \ref{fig:polyhedralCell-FVM}. For
the polyhedron in Figure \ref{fig:polyhedralCell-FVM}, the cell index in the
list of mesh cells is lower than the cell index of its neighbour accross the
face $f$. This makes the polyhedron the \emph{owner} of the face $f$, and makes
the ordering of the indices of the face $f$ such that the face area normal vector
$\vec{S}_f$ points outwards from the owner polyhedron, and into the neighbor polyhedron.
Such ordering allows for efficient (unique) computation of the volumetric
fluxes used by unstructured FVM. 

An algebraic approach to solving Equation \ref{equation:alpha-discrete}
utilizes bounded higher-order interface capturing (compressive) advection 
schemes on unstructured meshes to determine the face-centered volume fraction 
value $\alpha_f$. This practice often results in an artificial smearing of 
the volume fraction field as well as artificial deformation of the interface. 

The geometrical VoF algorithm solves Equation \ref{equation:alpha-discrete} by
computing the fluxed phase volume: 
\begin{equation}
    V_{f,\alpha} = \alpha_{f} \| \U_f \cdot \vec{S} \| \Delta t = \alpha_f \| F_f \| \Delta t
    \label{equation:vol-phase-flux}
\end{equation}
using geometrical operations, where $F_f$ is the volumetric flux across the face
$f$ with an outward pointing surface area normal vector $\vec{S}$ 
($\vec{S}_f$ for a computational cell formally owning the face $f$ -- cf.\ Figure 
\ref{fig:polyhedralCell-FVM}), which satisfies the discrete divergence free condition: 
\begin{equation}
    \sum_f F_f = 0.
    \label{eq:discretedivergence}
\end{equation}
Without the indirect addressing of the unstructured mesh, \eqref{eq:discretedivergence} 
would be computed on a per-cell basis, repeating the flux calculation twice per cell 
face. To avoid this, the flux is computed using the oriented face area normal vector 
$\vec{S}_f$ and the owner-neighbor addressing of the unstructured mesh: 
\begin{equation}
   \sum_f \alpha_f \vec{U_f}\cdot\vec{S} = \sum_{\text{owner}} \alpha_f \vec{U_f}\cdot\vec{S}_f 
     - \sum_{\text{neighbor}} \alpha_f \vec{U_f}\cdot\vec{S}_f, 
   \label{equation:own-nei}
\end{equation}
thus computing the same flux once per cell and adding the flux contribution to
the owner cell (outward directed $\vec{S}_f$) and deducting the flux
contribution from the neighbor cell (inward directed $\vec{S}_f$). 

Because of the geometrical multidimensionality of the directionally
unsplit advection algorithm, the artificial deformation of the interface which
is transported in the \emph{skew direction}\footnote{Skew direction in the mesh
is the direction of the point or edge neighbor cells.} is significantly
reduced, compared to fluxed-based methods based on algebraic differencing schemes. 
Such geometrical way of computing the fluxed phase volume also ensures the 
interface geometry to remain located within a single layer of cells throughout 
the simulation which is consistent with the discrete counter-part of the underlying 
sharp interface model. 

The geometrical VoF method relies on geometrical information of the interface. 
In our work, we have used a standard approach for the geometrical VoF
algorithm: a planar geometry of the interface is placed within an
\emph{interface cell} and the geometrical evolution of the volume fraction
field is performed by two subsequent steps, namely the \emph{interface
reconstruction} and the \emph{volume fraction advection} step. This algorithm
is commonly known as Piecewise Linear Interface Calculation (PLIC) algorithm
\citep{RiderReconstructing1998}.

In the remainder we provide a detailed explanation of the supporting
geometrical operations and both geometrical steps of our geometrical VoF
method. 

\subsection{Basic geometrical operations}

Two geometrical operations are the basis for both the reconstruction and the
advection step: the intersection between a plane and a polyhedron, and the
intersection between two polyhedrons. These operations belong to the field of
\emph{Computational Geometry} \citep{SchneiderGTC, DeBerg2008CG} and are
usually sources of difficulties when it comes to calculations that support
geometrical advection of the volume fraction in three dimensions. The reason
for this lies in the fact that the polyhedrons involved in the intersection
operations become non-convex due to the local complexity of the discrete
velocity field. When the full Lagragian discrete flow map is used to compute
the flux polyhedrons, complex flow fields may introduce non planarity of the
swept faces. As reported by \citet{AhnShashkov2009}, in case of a full
Lagrangian backtracking of the whole mesh, and subsequent re-mapping onto the
Eulerian static mesh, the faces of the swept cells may be non-planar, and the
resulting polyhedrons non-convex. If this occurs, a simplification of the swept
mesh is performed that enforces geometrical operations on convex polyhedrons
for the sake of simplicity and robustness of the overall algorithm.

Intersection between a plane and a convex polyhedron is calculated using the
\emph{polyhedron clipping and capping algorithm}, using a slightly modified
version of the algorithm described by \citet{AhnShashkovGeometricAlg2008}. As 
the method's name suggests, each face of the polyhedron is visited and 
\emph{clipped} by a plane, marking the new vertices that will at the end be 
added as the polyhedron \emph{cap} face. Once all the faces are clipped, the 
capping algorithm will append the final cap face making sure there are no 
repeated points. The polyhedron faces are defined to be oriented counter 
clockwise, when viewed from outside of the polyhedron. Special cases may 
appear during the geometrical operation, which may result in an intersection 
being represented as: void, point, edge, face or the complete polyhedron. 
All these cases are internally and automatically dealt with within the 
clipping part of the algorithm implementation.

\begin{figure}
    \centering
    \def\svgwidth{0.5\columnwidth}
       {\footnotesize
        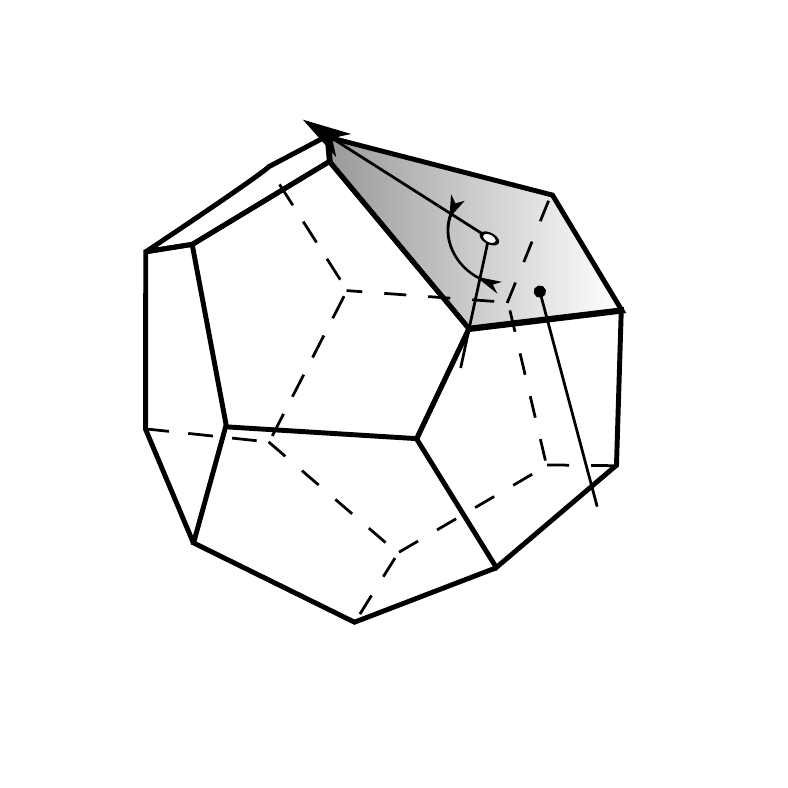
       }
    \caption{Cutting (clippig and capping) of a polyhedral cell}
    \label{fig:cap}
\end{figure}

The final cap face of the polyhedron is shown in Figure \ref{fig:cap}. The
order of the cap vertices $p_{cp}$ could be computed using the connectivity of
the clipped polyhedron: cap vertices belong to edges that have only a single
face attached to them after the clipping part of the algorithm is finished. In
Figure \ref{fig:cap}, $cp$ marks the cap point and $p_{c}$ is the center
point of the cap. However, we have sorted the points based on the angle the
vertex makes with respect to the cap center axis, given by the first cap point
candidate. Since we are dealing with polyhedra that have small number of faces
(finite volume cells usually have a small number of faces), an angle based
insertion sort is less complicated to implement and is efficient when compared
to constructing the face connectivity information to locate cap edges of the
polyhedron using a graph representation of the polyhedron. In order to find the
cap edges using edge-face connectivity, an edge based graph representation of
the polyhedron would be necessary. Once the clipping and capping of the
polyhedron is performed, the intersection between the polyhedron and a plane is
complete. 

The intersection between two convex polyhedrons can be built by applying the
polyhedron clipping and capping algorithm for each face of a polyhedron. The
same algorithm is used for calculating the intersection between a plane and a
cell, additionally ensuring the face normal is directed outwards from the
polyhedron created from the mesh cell in OpenFOAM. 

\subsection{Interface reconstruction}

We have chosen a first order accurate reconstruction algorithm similar to the
one of \citet{ParkerYoungsGradient} in which the gradient of the volume
fraction field determines the orientation of the interface plane with a
difference in the accurate and parallelized method of gradient calculation on
arbitrary unstructured meshes. This algorithm is first order convergent
regarding the geometrical reconstruction error, i.e. it cannot reconstruct
planar interfaces exactly. The absolute accuracy in determining the interface
orientation comes from the connectivity provided by the cell stencil on which
the volume fraction gradient is computed.
accurate gradient calculation method on a structured mesh is the Mixed
Youngs-Centered (MYC) gradient calculation method, developed by
\citet{Aulisa2007}. Using a wider stencil to increase the accuracy in
computing the interface normal introduces numerical smoothing of the interface
curvature, as reported by \citet{KotheUnstructured1996}. This is a common issue
of the  geometrical VoF algorithm, however, since the reconstruction error
reduces with the increase of mesh density, the application of local dynamic AMR
significantly reduces both the reconstruction errors and the need for a
reconstruction algorithm of higher order. Additionally, the overall
computational overhead of the solution algorithm is significantly reduced by
reducing the number of cells compared to a static uniformly refined mesh. 

The overall reconstruction algorithm is described by Algorithm
\ref{alg:reconstruction}, defined by two key steps: gradient
calculation and interface positioning. 

\begin{algorithm}[H]
    \centering
    \caption{Reconstruction algorithm}
    \label{alg:reconstruction}

    {\small
      \begin{algorithmic}
       \State compute the gradient of the volume fraction
       \For{cells}
         \If{cell is an interface cell}
             \State initialize the iterative positioning
             \While{interface not positioned}
                 \State secant method: update position
                 \If{secant diverging (two arguments outside search interval)}
                    \While{interface not positioned}
                     \State bisection method: update position 
                    \EndWhile
                 \EndIf
                 \State update secant divergence test (store last argument)
             \EndWhile
         \EndIf
        \EndFor
      \end{algorithmic}
    }
\end{algorithm}

\subsubsection{Gradient calculation}

Calculating the volume fraction gradient to second order of convergence on an
unstructured mesh has been investigated mostly for single phase simulations
(e.g. \citet{lsq-revisiting}) and for volume rendering algorithms in the field
of Computer Graphics (e.g. \citet{IEEE-gradComparison}). When it comes to two-phase 
flows on unstructured mesh, an overview of the possible gradient schemes
is provided by \citet{KotheUnstructured1996}.
\citet{AhnShashkovMultiMaterialRec2007} have used a least squares approach for
gradient calculation on an unstructured mesh for comparing the results with
their MoF reconstruction. 

A steep change of the volume fraction field across the fluid interface within a
single cell layer introduces difficulties in obtaining an accurate gradient
calculation on unstructured meshes. Computation of the gradient on unstructured
meshes based on the approximation stemming from the Gauss divergence theorem is
of second order convergence, however, this gradient scheme does not take into
account point cell neighbors. Therefore, the absolute accuracy of such schemes
deteriorates on fields with large jumps in values, across a narrow band of
cells. In the case of our geometrical VoF algorithm, the reduced absolute
accuracy of the standard Gauss gradient is observed in the form of
instabilities of the reconstructed interface geometry. The instabilities
persist regardless of the choice of the interpolation scheme, since the
interpolation stencil remains unchanged. This is especially true for
tetrahedral and hexahedral unstructured meshes. On polyhedral meshes, each cell
is connected to all neighbor cells by sharing a common face. The face-based
gradient calculation on polyhedral meshes is thus more accurate.

We have used a \emph{Node Averaged Gauss} (NAG) gradient scheme to compute the
orientation of the interface normal. The NAG gradient relies on the
connectivity which is already provided by the unstructured mesh. However, the
increase in absolute accuracy comes from successive interpolation of cell
centered values to mesh points, and averaging of the point values at face
centers. The value of the field is interpolated from cell centers to cell
points using an Inverse Distance Weighting (IDW) interpolation:
\begin{equation}
    \phi_p = \sum_{pc} w_{pc} \phi_{pc}
    \label{eq:idwpointphi}
\end{equation}
where $pc$ marks cells $c$ surrounding the mesh point $p$ (point-cells), 
$\phi_p$ is the field value which maps to mesh points, and $w_{pc}$ is the
inverse distance weight:
\begin{equation}
    w_{pc} = \frac{\frac{1}{||\x_p - \x_{pc}||}}{\sum_{\widetilde{pc}} \frac{1}{||\x_p - \x_{pc}||}},
\end{equation}
between the center of the point-cell $pc$ and the mesh point $p$. 

From the mesh point values, the face-centered values are calculated using
\emph{area weighted averaging}. As a first step, the center of the face is
calculated as the algebraic average of the face points, and the initial
face-centered value $\phi_{f}'$ is computed by algebraically averaging the
point values: 
\begin{equation} \phi_{f}' = \frac{1}{N_{fp}} \sum_{fp} \phi_{fp},
\end{equation}
where $\phi_{f}$ is the \emph{initial face centered value}, $fp$ is a set of
mesh points that comprise a face $f$ (face-points) and ${N_{fp}}$ is the number
of face-points. In a second step, the face is decomposed into triangles using
the face center and points of each face edge ($fp$, $fp+1$). Triangle area is
computed using the vector product its edge vectors, and is then used as a
weighting factor: the final face centered value is then computed as an area
weighted average of the averaged triangle values: 
\begin{equation} \phi_f = \sum_t w_t \phi_t, \end{equation}
where $w_t$ is the area weight of the triangle:
\begin{equation} w_t = \frac{A_t}{\sum_t A_t}, \end{equation}
and $\phi_t$ is the average value of the triangle: 
\begin{equation} \phi_t = \frac{1}{3} (\phi_f' + \phi_{fp} + \phi_{fp+1}).
\end{equation}
As a final step in computing the cell centered gradient, the face centered
field is used to calculate the gradient using the Gauss divergence theorem in
the gradient form: 
\begin{equation} 
\grad{\phi}_c \approx \frac{1}{V_c} \sum_f \phi_f \vec{S}_f. 
\end{equation}
This algorithm accounts for the volume fraction values stored in \emph{skew
cells} (cells which are connected with the cell in question through a point or
an edge) with their information to be transferred to the interface cell
indirectly, without the explicit computation of the additional connectivity
required by the wide cell stencil on the unstructured mesh. 

The area weighted averaging takes into account the contribution to the final
face-centered value by taking into account the shape of the face. An
alternative calculation of the NAG gradient could involve IDW interpolation
applied to compute the face centered values instead of the area averaging
procedure. This alternative has not been tested since the above averaging 
approach has been used to produce very accurate as well as convergent 
interface reconstruction results.

The \emph{generalized least squares gradient} (generalized LSQ) that supports
arbitrary unstructured meshes as described by \citet{lsq-revisiting} has also
been implemented. For the generalized LSQ gradient, the connectivity stencil
storing the access information to cell point neighbors needs to be
constructed. For a linear least squares gradient, the field is approximated
using the linear part of the Taylor series, i.e. 
\begin{equation}
    \phi(\x_{k}) \approx  \phi(\x_{c}) + \grad{\phi}(\x_{c}) (\x_{k} - \x_{c}),
\end{equation}
where $c$ and $k$ are the indirect access indices of two cell centers on an
unstructured mesh. Since a wider cell stencil on a three-dimensional mesh
implies a larger number of cell centers than three, a gradient of a scalar
field $\phi_c$ is computed by minimizing the weighted gradient error for each cell $c$:
\begin{equation}
    E_c = \sum_k w_{ck}^2 E_{ck}^2,   
    \label{eqn:generalizedLSQminimize}
\end{equation}
where the error is defined as (\citep[]{lsq-revisiting})
\begin{equation}
    E_{ck} = -[\phi(\x_{c}) - \phi(\x_{k})] + \grad{\phi}(\x_c) (\x_c - \x_k),
\end{equation}
and the $w_{ck}$ is the inversed distance weight. The minimization of the error
$E_c$ for a scalar field results with a solution of a  algebraic equation
system for each cell of dimensions $3\times3$, with the dimensions defined by
the three components of the gradient vector. 

\subsubsection{Interface positioning}

An iterative approach to interface positioning
(\citet{RiderReconstructing1998}) using the super-linear convergent Brent's
algorithm (\citet{BrentAlgorithm}) or an analytical approach
(\citet{ScardovelliAnalytical2000}) can be chosen for interface positioning.
\citet{AhnShashkovMultiMaterialRec2007} have shown that the interface
positioning may be accomplished in a simpler way, using a combination of the
secant and bisection iteration methods. The bisection method has guaranteed
linear convergence, and the secant method has super-linear convergence, but may
diverge. 

\begin{figure}
    \centering
    \def\svgwidth{0.5\columnwidth}
    {\footnotesize
        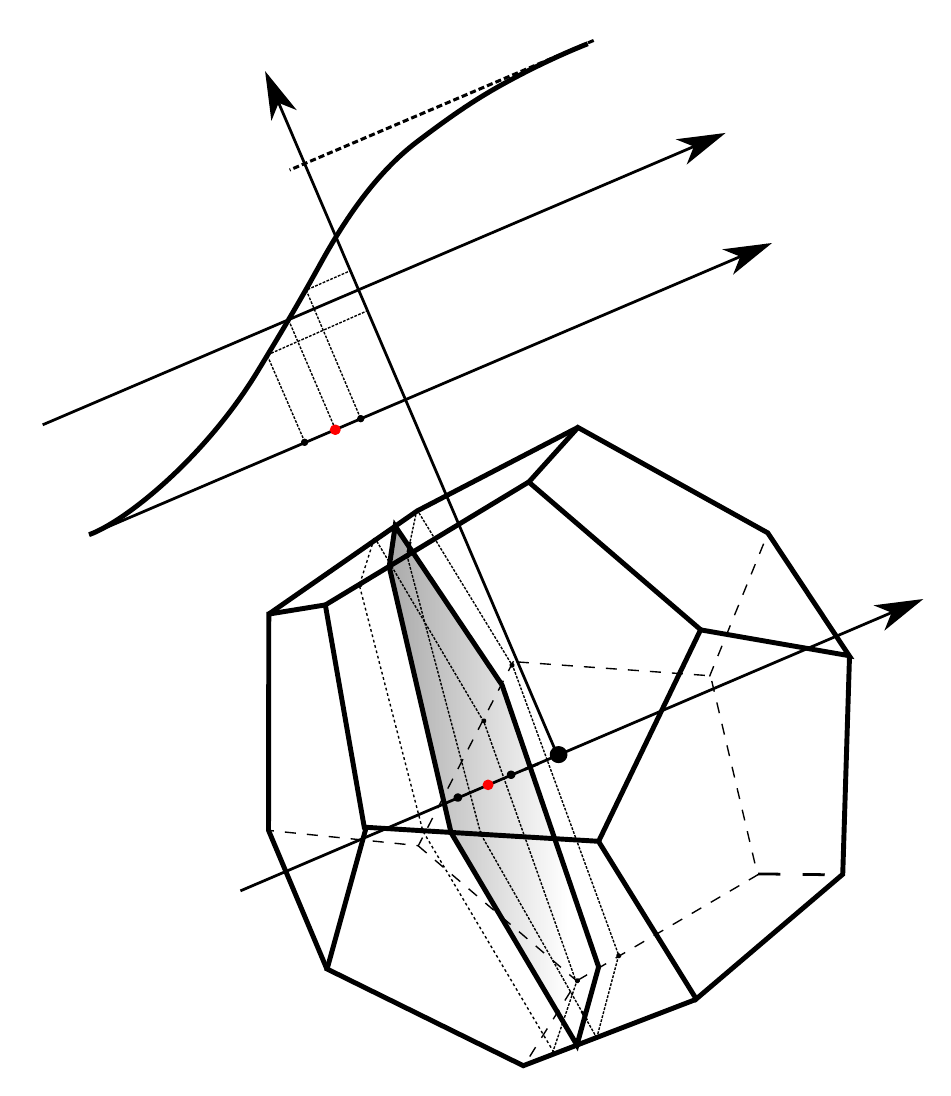
    }
    \caption{Geometric Reconstruction: Interface positioning}
    \label{fig:positioning}
\end{figure}

Figure \ref{fig:positioning} shows the details of our interface positioning
algorithm. A characteristic fill level function is shown which corresponds to
the increase of the volume fraction value as the interface is moved along the
orientation axis given by the gradient of the volume fraction. Because of the
monotonicity of the fill level function we know that the fill level function
has a zero value within the search interval. However, divergence of the secant
method may happen when the iterative procedure reaches the border of the
interval since the fill level function has a \emph{vanishing slope} at the
interval borders ($-\alpha$ or $1 - \alpha$), so the iterative procedure
switches to the bisection method in this case. 

\subsection{Volume fraction advection}

\begin{figure}
   \begin{subfigure}[b]{0.45\columnwidth}
       \centering
       \def\svgwidth{\columnwidth}
       {\footnotesize
            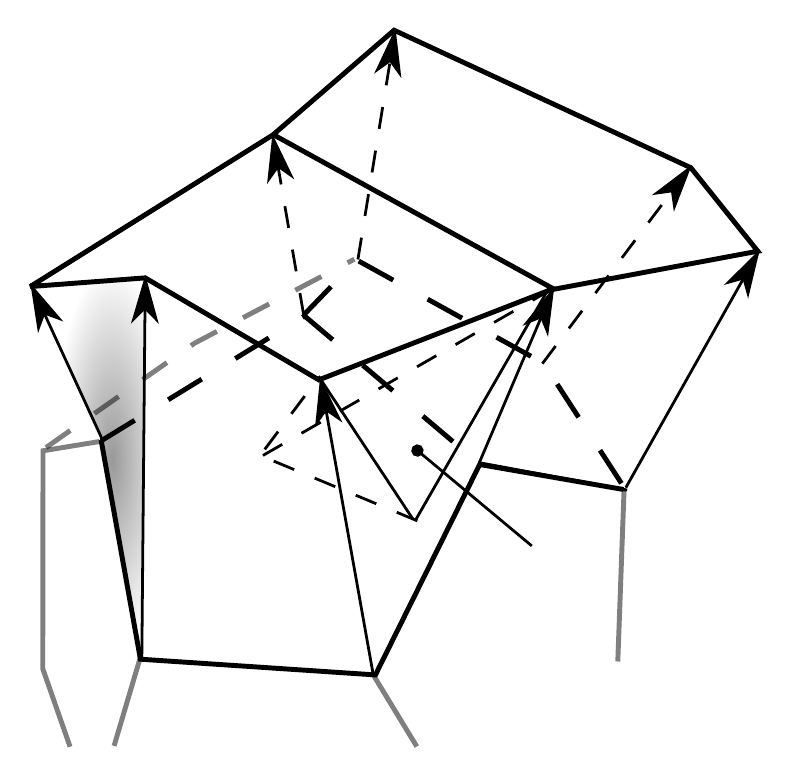
       }
       \caption{Sweeping faces with point velocities produces non-overlapping swept polyhedra} 
       \label{fig:sweptPolyhedron:sweeping}
   \end{subfigure}
   \begin{subfigure}[b]{0.45\columnwidth}
       \centering
       \def\svgwidth{\columnwidth}
       {\footnotesize
            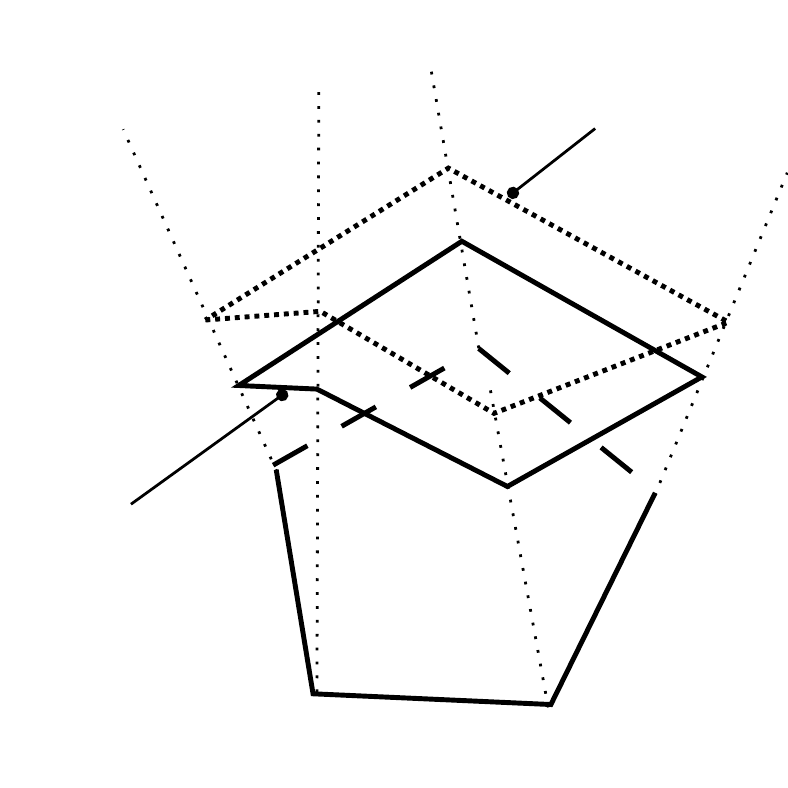
       }
       \caption{Iterative correction of the swept polyhedron to match the volume given by the volumetric flux}
       \label{fig:sweptPolyhedron:correcting}
   \end{subfigure}
\end{figure}

Advection of the volume fraction is based on a geometrical computation of the
\emph{fluxed phase volume} $V_{\alpha}$ on cell faces. This is achieved by
\emph{sweeping} (backtracking) the face along the streamline vectors given by
the interpolated velocity field at the mesh points: 
\begin{equation} 
\vec{x}'_{fp} = \vec{x}_{fp} - \vec{U}_{fp}\Delta t,
\label{eq:sweep} 
\end{equation}
where $\vec{x}_{fp}$ is the face-point (a mesh point addressed via the cell
face index list).  The swept face, together with the original face, construct
together a so-called \emph{congruent} polyhedron, which may have non-planar faces and
may even be non-convex. The advantage of using unique point velocities lies in
the fact that the \emph{swept polyhedra do not overlap at all}, as shown in
Figure \ref{fig:sweptPolyhedron:sweeping}. However, it has been reported in the
literature \citep{LiovicCVTNA2006} that using the full discrete flow map may
result in highly distorted polyhedra when complex flow fields are present,
introducing errors in mass conservation and numerical boundedness of the volume
fraction field. To the contrary, as we show later in the results part, below the 
expected Courant number limit ($Co<1$) our development of the numerical flow 
solution remains mass conservative and bounded within 
$[0,1]$ using point velocities, even for complex flow fields and both on meshes 
with densities used for standard test cases and on dynamically refined meshes 
using local dynamic AMR. %
We have observed an increased accuracy and robustness of the advection algorithm 
because of the tetrahedral decomposition applied to the computation of the volume 
of the swept polyhedron. %
Analytical expressions for computing the volume of the flux polyhedron rely
exclusively on the convexity of the polyhedron. Using unique point velocities
to sweep the face does introduce some non-planarity of the faces of the swept
as shown for an example shaded non-planar face in Figure
\ref{fig:sweptPolyhedron:sweeping}. However, the distortion of the polyhedron
will always be small enough, such that the tetrahedral decomposition removes the
face non-planarity and delivers accurate results for the volume of the swept
polyhedron $S_f$ with non-planar faces, which allows for an efficient iterative
flux correction procedure that completely avoids topological operations on the
swept polyhedron. 

Figure \ref{fig:sweptPolyhedron:sweeping} shows the construction of the swept
polyhedron: face point velocities $\U_{fp}$ are interpolated using 
\eqref{eq:idwpointphi} and multiplied by $-\Delta t$ to track the
face back (Equation \ref{eq:sweep}). Tetrahedral decomposition is then used to
compute the volume of the volumetric flux per each face $f$ and the polyhedron
face is swept along the stream vectors iteratively (Figure
\ref{fig:sweptPolyhedron:correcting}) until the volume of the swept polyhedron
becomes equal to the volume given by the volumetric flux $F_f$: 
\begin{equation}
    V_{f,F} = \| F_f \| \Delta t,
    \label{eq:VF}
\end{equation}
converging into the final \emph{flux polyhedron} used to compute the volume of
the fluxed phase accross face $f$.
The velocity field obtained at the face points using inversed distance
interpolation is not divergence free in the discrete sense, thus a divergence 
correction becomes necessary. For this, we define the volume of the face-swept 
flux polyhedron:
\begin{equation}
    V_{f,S} = \sum_{f,T} V_{f,T}, 
\end{equation}
where $V_{f,T}$ is the volume of a tetrahedron produced as a result of
tetrahedral decomposition of the swept-face polyhedron $S_f$ generated for each
face $f$ of the mesh (Figure \ref{fig:sweptPolyhedron:sweeping}).  Using the volume
of the decomposed face-swept flux polyhedron, the discrete swept volumetric
flux can be computed as
\begin{equation}
    F_{f,S} = \frac{V_{f,S}}{\Delta t}.
\end{equation}
As already noted, this volumetric flux is not expexted to satisfy the discrete 
divergence free condition \eqref{eq:discretedivergence} necessary for the mass 
(volume) conservation of a pseudo-staggered solution algorithm, i.e.\
\begin{equation}
    \sum_f {F_{f,S}} \sgn(F_f) \ne 0,
\end{equation}
where $\sgn(F_f)$ is the sign function of the volumetric flux: 
\begin{equation} 
   \sgn(F_f) = \begin{cases} 
                    +1 , F_f > 0 \\
                    -1 , F_f < 0  \cdot  
              \end{cases},
\end{equation}
%
%
which needs to be computed once per face since the volume of the swept-face polyhedron 
will always be positive. Subsequently a \emph{divergence correction} of the swept 
polyhedrons is to be performed to ensure mass conservation. %
This practice has proven sufficient for the computation of the fluxed phase volumes, 
since the point velocity field is then only needed to provide the direction for face 
sweeping (streamline direction).%

In literature, there are three approaches commonly used for the divergence
correction: a scalar coefficient correction \citep{LiovicCVTNA2006}, analytical
correction \citep{HernandezLopezPart1Advection2008} and a parametric correction
\citep{MencingerZunAdaptive2011}. We have developed a novel efficient and 
iterative approach: %
the \emph{swept polyhedron} computed by sweeping the face is corrected iteratively 
until its volume is equal to the volume given by the fluxed volume defined by the 
scalar volumetric flux (Equation \ref{eq:VF}).
The super-linear convergent secant method is applied to \emph{move the end points
of the flux polyhedron} until its volume corresponds to the volumetric flux volume. 
This algorithm does not involve topological changes of the flux polyhedron and the 
method converges in a very small number of steps (3-5 on average, evaluated for 
the shear flow test case) to machine tolerance since the initial volume of the flux 
polyhedron is in magnitude near to the magnitude of the volumetric flux, and the 
target function is a fill level function with a monotonic increase. Avoiding 
topological operations in the correction of the flux polyhedron increases the speed 
of the correction significantly. We have tested the algorithm for different shapes 
of swept faces, applying random perturbations in the orientation of swept edges 
and have found the correction algorithm to be robust and accurate. Even with 
random perturbations of swept point coordinates in magnitudes up to 20\% of the 
characteristic length of the face (Figure \ref{fig:sweptPolyhedron:sweeping}), 
the calculation of the volume magnitude remains correct to machine tolerance, 
allowing for the iterative correction algorithm to converge very quickly. 


\begin{figure}
       \centering
       \def\svgwidth{\columnwidth}
       {\footnotesize
           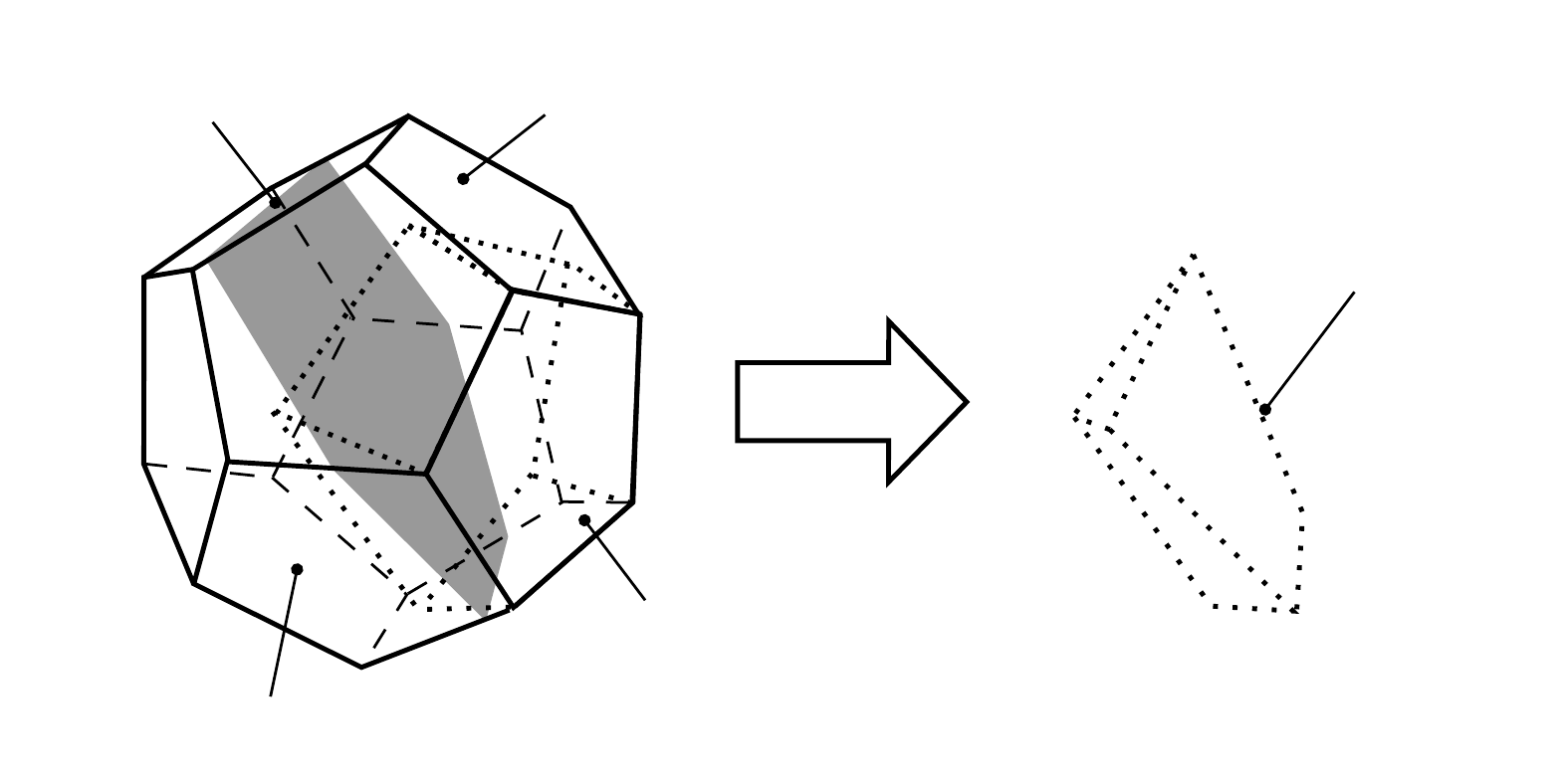
       }

    \caption{Computing the volume of the fluxed phase using intersections.} 
    \label{fig:volPhaseFlux}
\end{figure}

After the swept polyhedrons have been corrected, the contributions of the volumetric
phase flux coming from the surrounding cells needs to be computed. Obtaining a
flux contribution using geometrical intersections is shown schematically in
Figure \ref{fig:volPhaseFlux} for a single face. The corrected swept 
polyhedron $S_f$ (dotted lines) is initially clipped with each cell of the 
advection stencil:
\begin{equation}
   S'_f = S_f \cap_s C_s,  
\end{equation}
where $C_s$ is the cell of the chosen flux stencil $s$. The flux stencil $s$
needs to incorporate point and edge face neighbors, otherwise those
contributions to the total fluxed phase volume will be missing and interface
motion will be artificially distorted in those directions, as is the case
for flux-based methods naively utilizing algebraic advection schemes on 
unstructured hexahedral meshes. In case the candidate cell is an interface 
cell, a subsequent intersection is performed with the positive halfspace 
$I^+$ defined by the PLIC plane (grey colored polygon):
\begin{equation}
    S''_f = S'_f \cap I^+.
\end{equation}
This is the final face contribution to the total fluxed phase volume as illustrated
in Figure \ref{fig:volPhaseFlux}. The total value of the volumetric flux is
then computed as the sum of the fluxed phase volume contributions calculated for
all cells of the face flux stencil. The calculation of the total fluxed phase
volume across a face is performed for an \emph{outflow cell} only using 
\eqref{equation:vol-phase-flux}, resulting in a final decrease of the volume 
fraction for this cell, and the corresponding increase for the \emph{inflow cell}. 

We use a \emph{narrow band} of interface cells for the advection of the volume
fraction field. 
Since geometrical intersections are needed for the geometrical reconstruction and
advection steps, localizing the computation to a narrow band of cells strongly
increases the efficiency of the overall algorithm. In case of a complex
velocity field and a naive advection operating on the whole domain,
\emph{wisps} appear in the bulk of both phases. Hence, locating and removing
wisps is greatly simplified for a \emph{narrow band} advection since their
initial location is always in the bulk cells that are direct neighbors to
interface cells. To remove the wisps, a redistribution algorithm of
\citet{HarvieFletcherStream2000} has been implemented, with applied
modifications that are necessary for fast computation of large stencils on
arbitrary unstructured meshes. Both void and phase wisps are redistributed
conservatively to the surrounding interface cells, under the constraint that
the removal of mass or its addition is not allowed to cause further creation of
over/undershoots or wisps. 

The narrow cell band is constructed using neighborhood cell information and the
volume fraction field, i.e. the algorithmic task for the advection is: advect
only in interface cells and their first \emph{point neighbors}. To avoid
computing additional point-cell neighbor connectivity which would then be
updated during the evolution of the interface, we have re-used the gradient of
the volume fraction field as an indicator for the bulk cells that are not
involved in the advection. This is schematically shown in Figure
\ref{fig:parallel:gradZeroCondition}.  A bulk cell is  characterized as a cell
for which the gradient of the volume fraction is zero.

Since only the outflow from cells is calculated, the change of the volume
fraction is not applied for cells that lie in the bulk of the phases.  The
fluxed phase volume is computed also for cell faces when the outflow is
present for the bulk cell. However, in this case the update of the volume
fraction field is not executed for the bulk cell. The re-use of the gradient
field as a criterion for constructing the narrow band of cells involved in the
advection significantly increases the efficiency of the method, since the
gradient is already computed as a requirement of the reconstruction algorithm.

\subsection{Parallel implementation}

The parallel implementation of the reconstruction algorithm is simplified when
the NAG gradient is utilized for the computation of the interface normal, since
no additional communication structures need to be prepared and communicated
over the process boundary. In the case of the generalized LSQ algorithm,
computing the additional connectivity involves multiple issues: parallel
communication involving more complex stencil structures, update of the stencil
structures when topological changes are applied to the mesh (AMR), and caching
of the additional connectivity to increase algorithm efficiency. Since the NAG
gradient has provided sufficiently accurate results, we have left the parallel
implementation of the generalized LSQ gradient as future work if proved to be
necessary.

\begin{figure}
   \begin{subfigure}[b]{0.45\columnwidth}
       \centering
       \def\svgwidth{0.6\columnwidth}
       {\footnotesize
           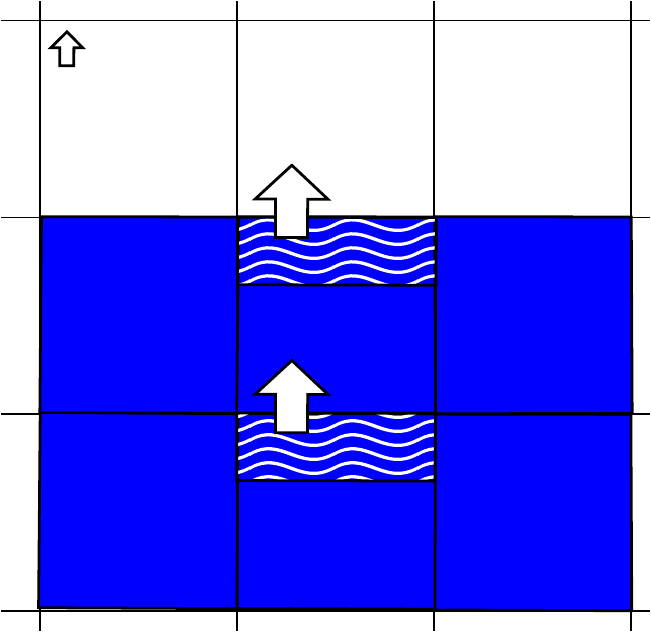
       }
       \caption{Computing a narrow band of cells} 
       \label{fig:parallel:gradZeroCondition}
   \end{subfigure}
   \begin{subfigure}[b]{0.45\columnwidth}
       \centering
       \def\svgwidth{0.6\columnwidth}
       {\footnotesize
           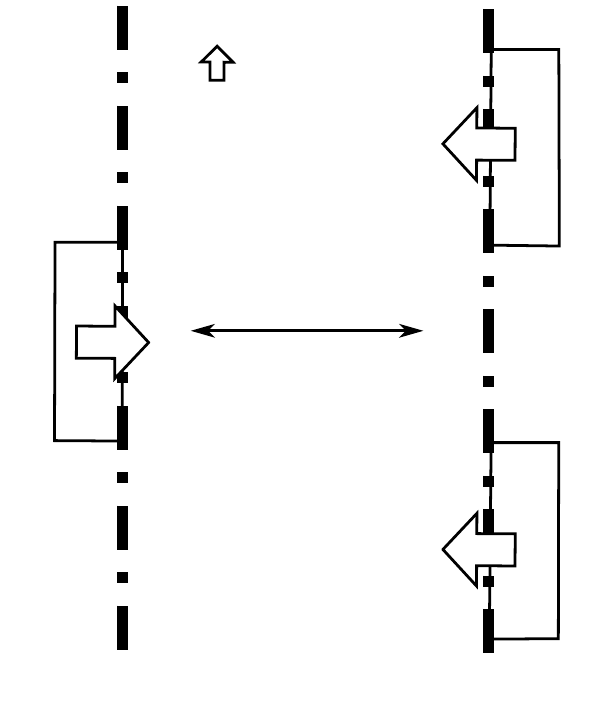
       }   
       \caption{Communicating the outflow flux fields}
       \label{fig:parallel:communication}
   \end{subfigure}

    \caption{Parallel implementation: computing the fluxed phase volume}    
    \label{fig:parallel:sphereFine}
\end{figure}

Using the gradient to determine the bulk cells simplifies the parallel
implementation of the advection algorithm as shown schematically on Figure
\ref{fig:parallel:gradZeroCondition}. Additional connectivity required to
identify the cell candidates would otherwise be communicated across process
boundaries, as well as the geometrical information of the mesh cells. However,
when the volume fraction gradient is used to determine the narrow band of
cells, there is no need for additional connectivity. Since the outflow field is
computed for each separate domain, the boundary outflow fluxes need to be
exchanged (simultaneously) across process boundaries. The outflow from one
domain represents the inflow field for the other domain, as shown in Figure
\ref{fig:parallel:communication}. In order to avoid \emph{deadlocking}
resulting from simultaneous communication, the parallel code utilizes \emph{non
blocking} communications \citep{OpenMPI}.

\section{Geometrical mapping of the volume fraction field for local adaptive mesh refinement}
\label{section:geometrical-mapping}
Although the geometrical VoF algorithm supports arbitrary cell shapes, the mesh
topology dictates the concrete strategy and conceptual approach (i.e.\ type of
topological operations) for mesh refinement. We are applying local dynamic AMR
to increase the solution accuracy especially in the vicinity of interfaces, in
particular for the coupling between the geometrical algorithm and the flow
solution. The algorithm for local dynamic AMR in OpenFOAM is developed on the
basis of the unstructured \emph{hexahedral} mesh: it is based on a \emph{fully
unstructured mesh refinement} procedure. Fully unstructured mesh refinement has
the advantage to deal with flow domains of arbitrary geometrical complexity,
since the topological operations on the mesh are done directly, changing the
overall mesh connectivity information. However, since the connectivity
information is changed by the topological operation, the mesh data structures
need to be modified in order to store the new connectivity produced by the
refinement cycle.  

If the topological operation introduces a production of mesh data, the capacity
of the data structures needs to be extended. This involves re-allocating memory
and extending the capacity of the data structures, which sets a computational
price to the flexibility in dealing with arbitrary complex geometries. Even so,
the mesh refinement engine in OpenFOAM, together with the mapping of all
intensive properties, takes up only a small fraction of the overall time
involved in the flow solution.

The refinement cycle consists of two sub-cycles which may be executed,
depending on the refinement criteria: refinement and un-refinement sub-cycle.
If both sub-cycles are executed, the mesh information from the beginning of the
refinement cycle is lost in between the sub-cycle execution. Removal of
intermediate information makes it impossible to map the mesh modified by two
subsequent refinement cycles to the initial un-refined mesh to which the
interface geometry maps. The obstructed mapping renders the refinement
procedure unusable together with the geometrical transport algorithm. For this
purpose, we have extended the dynamic mesh capability to store the connectivity
information relative to the unmodified mesh state at the beginning of the
refinement cycle. This further enables the execution of a \emph{geometrical
mapping algorithm} for the volume fraction field described by Algorithm
\ref{alg:geometrical-mapping}. 

\begin{algorithm}[H]
    \centering
    \caption{Geometrical mapping algorithm for the volume fraction field}
    \label{alg:geometrical-mapping}

    {\small
      \begin{algorithmic}
        \State initialize the absolute cell map 
        \If{cells to be refined exist}
           \State mark mesh as refined
            \State refine mesh 
            \State absolute cell map = current cell map
        \EndIf
        \If{cells to be unrefined exist}
           \State mark mesh as unrefined 
           \State copy the absolute cell map 
           \State un-refine mesh 
        \EndIf
        \If {mesh is refined and unrefined}
            \For{current cell map} 
               \State absolute cell map [cell] = absolute cell map[current cell map[cell]]
            \EndFor
        \ElsIf{mesh is unrefined}
            \State absolute cell map = current cell map
        \EndIf
        \For{absolute cell map}
            \State intersection = intersect (interface planes [absolute cell map [cell]], cell) 
            \State volume fraction [cell] = intersection / cell volume 
        \EndFor
      \end{algorithmic}
    }
\end{algorithm}

The interested reader is directed to our forthcoming publication of the authors for details
regarding the description of the unstructured mesh topology in OpenFOAM and its
reflection onto the geometrical mapping algorithm for the volume fraction field
\citep*{MaricVoFoamMapping2013}.

\section{Results}
\label{section:results}
\subsection{Verification of the interface reconstruction}

For structured meshes, a standard approach to computing the error of the
interface reconstruction algorithm (i.e.\ reconstruction error) resorts to
computing the volume that lies between the piecewise planar reconstructed
interface geometry and an analytical function representing the true interface
(e.g.\ the surface of a sphere).  This \emph {integration approach} requires
sub-integration for each cell.  Hence, it is not straightforward to modify the
algorithm in order to support arbitrary cell shapes.  

On arbitrary unstructured meshes, in order to achieve a single layer of
interface cells for a pre-processed volume fraction field, we have adopted the
approach by \citet{AhnShashkov2009} that involves intersecting two unstructured
meshes: the \emph{base mesh} and the \emph{tool mesh}. The \emph{base mesh} is
the mesh used to discretise the flow domain, and the \emph{tool mesh} is used
as a tool for cutting the \emph{base mesh} in order to set the volume fraction
field. This preprocessing approach of intersecting meshes allows for
preprocessing of arbitrarily complex volume fraction fields. Since the tool
mesh is unstructured, it may be used to discretize the complex input geometry.
Moreover, local refinement can easily be applied near boundaries that are used
to describe the fluid interface: this significantly reduces the computational
time of the preprocessing. The quality of the tool mesh is of virtually no
importance, since it is only use to set the initial volume fraction fields

In order to compute the reconstruction error on meshes with arbitrary cell
shapes, the reconstruction error is defined in each cell as the \emph{volume of
symmetric difference} ($E_{vsd}$) between the reconstructed geometry
(\emph{interface polyhedron}) and the true geometry. The true interface
geometry is represented by a subset of cells of a tool mesh, i.e.\ a set of
tool mesh boundary faces represent the true discrete interface geometry. 

\begin{figure}
   \centering
   \def\svgwidth{0.6\columnwidth}
   {\footnotesize
       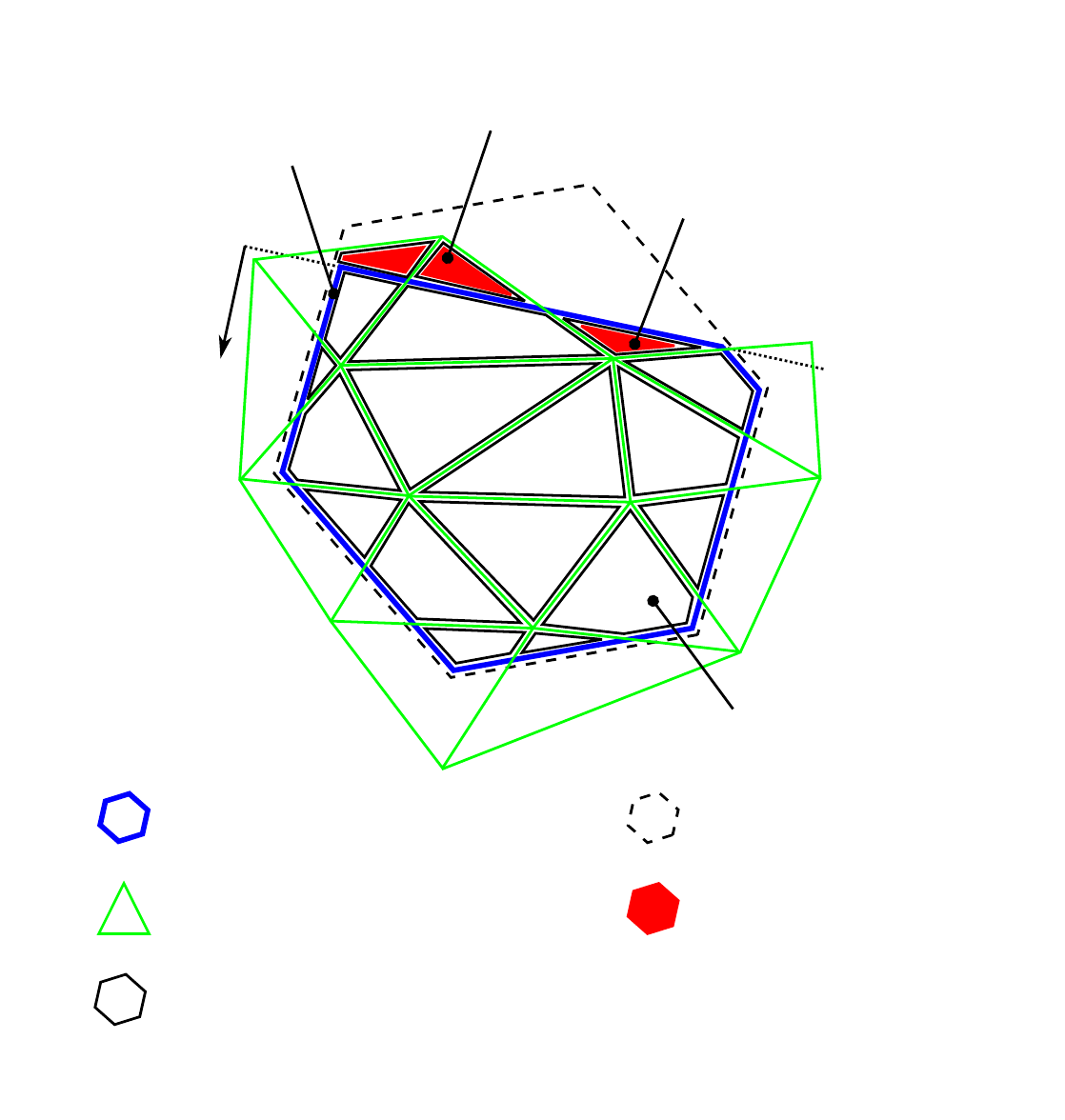
   }
   \caption{Volume of symmetric difference reconstruction error}
   \label{fig:volSymmDiff}
\end{figure}

The computation of the reconstruction error for a true interface geometry
defined by the tool mesh is performed cell-wise. Error calculation is
schematically described in two dimensions in Figure \ref{fig:volSymmDiff}. Each
mixed cell of the base mesh (dashed line) is intersected with a set of
polyhedrons of the tool mesh (green line). Each tool mesh polyhedron is then
clipped by the base mesh cell (black line). The reconstruction error (red
color) is separated into two contributions: \emph{positive half-space error}
($E_{vsd}^+$) and \emph{negative half-space error} ($E_{vsd}^-$), based on the
location of the error geometry with respect to the interface plane.

The set of tool mesh polyhedra is separated into two subsets based on the
interface plane orientation. The negative half-space error is the volume of
symmetric difference that lies in the negative closed half-space defined by the
interface plane. It is computed as an intersection between the set of the tool
mesh polyhedra clipped by the cell ($I_c$) and the negative half-space
($H(\vec{n})^-$), defined by the interface plane:
\begin{equation}
        E_{vsd}^{-} = || H(\vec{n})^- \cap I_{c} ||. 
\end{equation}
The positive half-space error is calculated as the algebraic difference between
the volume of the reconstructed (PLIC) polyhedron ($R_c$, blue line) and the
sum of the volumes of the clipped tool mesh polyhedra (black line): 
\begin{equation}
        E_{vsd}^+ = || R_c || - ||H(\vec{n})^+ \cap I_c||. 
\end{equation}
The total reconstruction error is computed as the sum of the positive
half-space error and the negative half-space error: 
\begin{equation}
        E_{vsd} = E_{vsd}^- + E_{vsd}^+.
\end{equation}
Note that when the \emph{integration approach} is used to determine the
reconstruction error the size of the sub-grid integration intervals will
determine the accuracy of the computed error (\cite{Aulisa2007}). The
\emph{mesh intersection approach} requires that the mesh density of the tool
mesh to be high enough in order to accurately describe the true interface
geometry. This is both visually observed and quantified for all validation
cases.   

\subsubsection{Reconstruction of a sphere} 

In this section the reconstruction results for a spherical interface on an
unstructured hexahedral, tetrahedral and polyhedral mesh are presented.  

\begin{figure}
	\centering
		\includegraphics{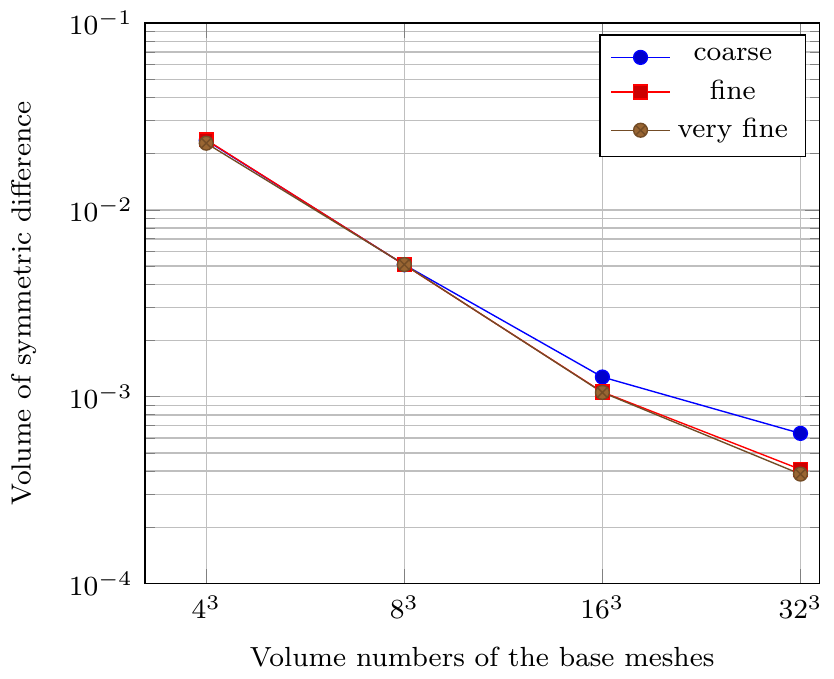}
    \caption{Reconstruction error for a sphere using the NAG gradient, on an unstructured hexahedral mesh} 
    \label{fig:recError:sphere:NAG}
\end{figure}

%
\begin{figure}
	\centering
		\includegraphics{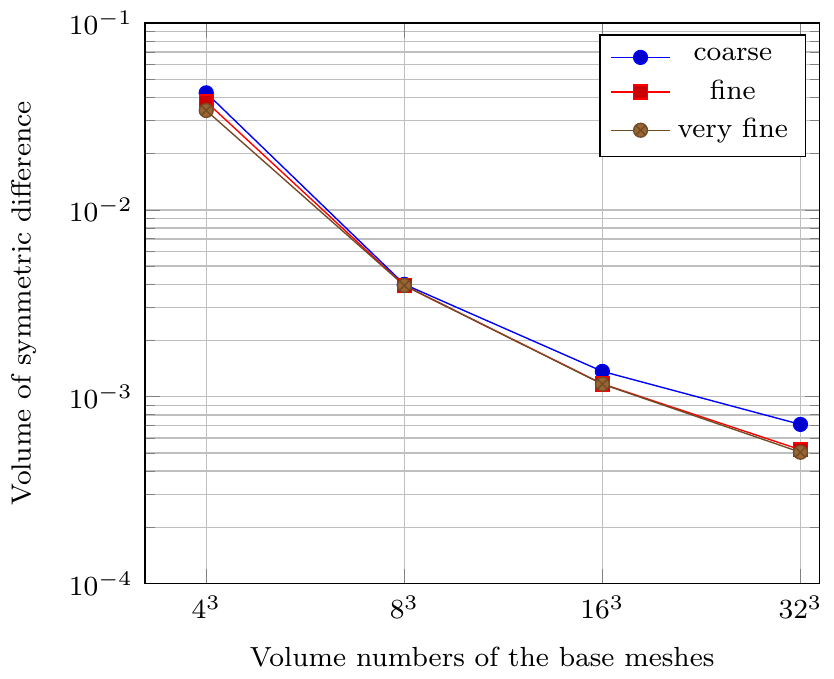}
    \caption{Reconstruction error for a sphere using the generalized LSQ gradient, on an unstructured hexahedral mesh}
    \label{fig:recError:sphere:generalizedLsq}
\end{figure}

Figures \ref{fig:recError:sphere:NAG} and
\ref{fig:recError:sphere:generalizedLsq} show the convergence diagrams of the
volume of symmetric difference reconstruction error on a box domain of
dimensions $[0,1]^3$ which is discretized with an \emph{unstructured hexahedral
mesh}. Although the discretization results in cuboid finite volumes, the mesh
connectivity is still unstructured. The first order Youngs algorithm exhibits
the behavior as described by \cite{AhnShashkovMultiMaterialRec2007}: on coarse
meshes the order of convergence of the error is higher and as the base
mesh density increases, the order reduces to around one. For the fine
tool mesh, the order of convergence as described by \cite{Aulisa2007} is
$2.22$ for the coarsest base mesh and $1.37$ for the finest base
mesh. 

\begin{figure}
   \begin{subfigure}[b]{0.45\columnwidth}
       \centering
       \includegraphics[width=\columnwidth]{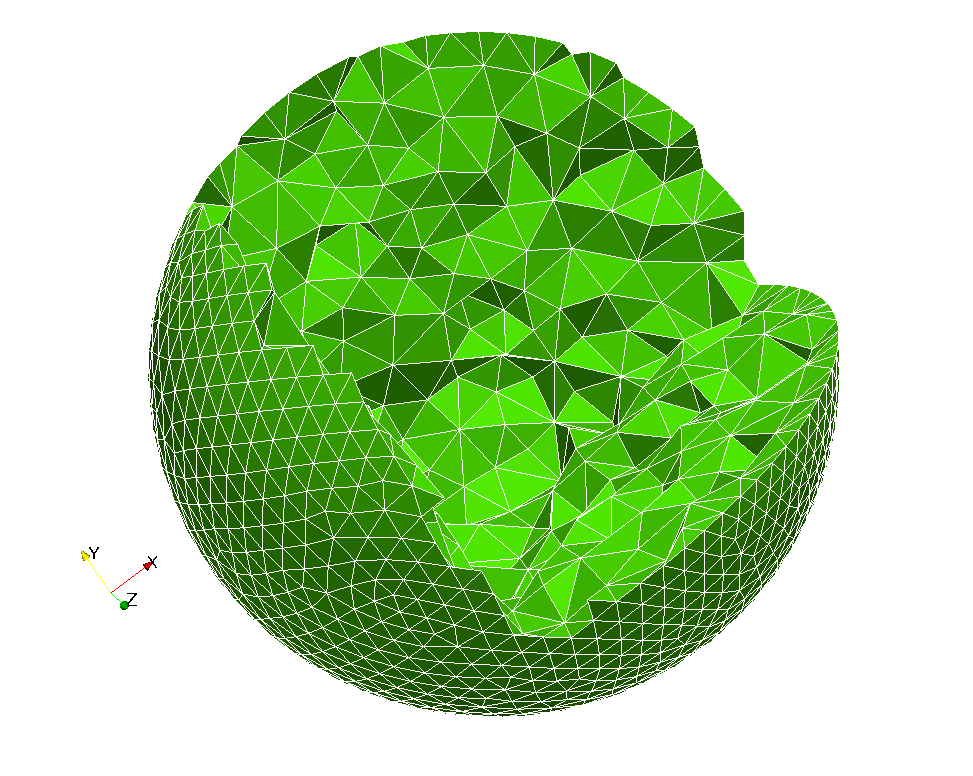}
       \label{fig:mesh:sphereFine:unrefined}
       \caption{Unrefined tool mesh}
   \end{subfigure}
   \begin{subfigure}[b]{0.45\columnwidth}
       \centering
       \includegraphics[width=\columnwidth]{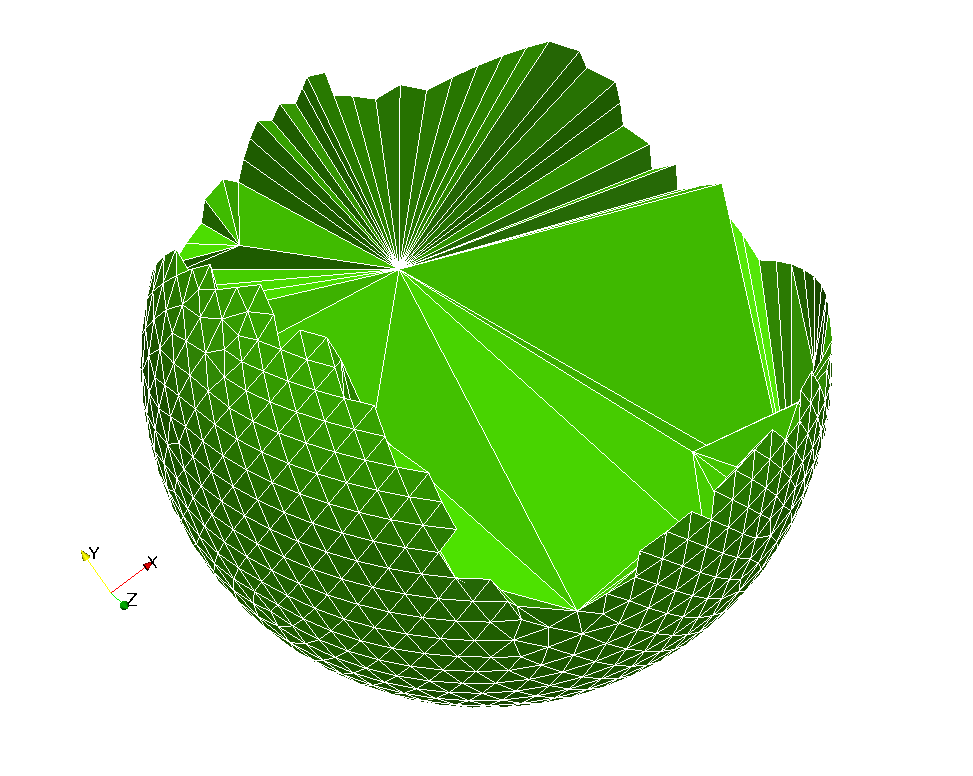}
       \caption{Refined tool mesh}
       \label{fig:mesh:sphereFine:refined}
   \end{subfigure}

    \caption{Tool meshes of a spherical interface}
    \label{fig:mesh:sphereFine}
    \label{fig:mesh:sphereFine:unrefinedRefined}
\end{figure}

\begin{figure}
	\centering
		\includegraphics{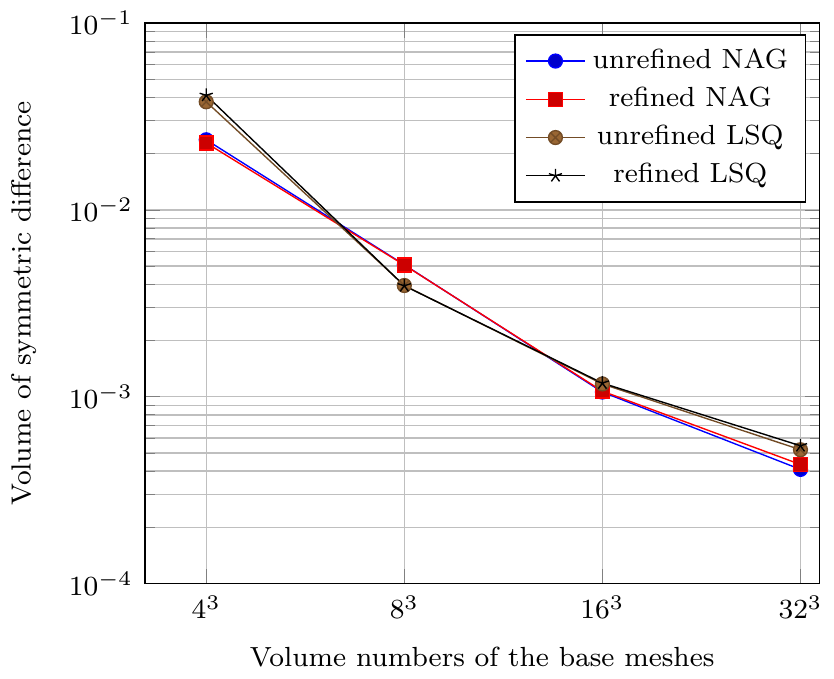}
    \caption{Comparison of the reconstruction error convergence for the refined and unrefined tool mesh for a sphere}
    \label{fig:recError:sphere:unrefinedRefined}
\end{figure}

Different mesh densities were tested for a spherical tool mesh. The
local refinement of the tool mesh near the interface makes the
total number of cells not important rather the number of faces used to approximate
the spherical surface is central. Figure \ref{fig:mesh:sphereFine} shows two \emph{tool
meshes} of a sphere, each with the same number of face elements, but different
number of volumes. Both meshes are cut such that the inner mesh is visible: the
refined mesh has much less volumes, and both meshes describe the spherical
surface with the same mesh density. The tool mesh does not need to be of
high quality since it is not a part of the solution process, i.e. it is used
only for the preprocessing of the volume fraction field. Figure
\ref{fig:recError:sphere:unrefinedRefined} shows the convergence for the
reconstructed spherical interface on an unstructured hexahedral mesh with two
sphere tool meshes, namely a refined sphere, and an unrefined sphere (cp.\ 
Figure \ref{fig:mesh:sphereFine:unrefinedRefined}). 

For the mesh convergence study we have used spheres discretized with $956$,
$3786$ and $16462$ faces\footnote{The meshes were created with the \emph{open
source} application \emph{SalomeCAD}. They can be easily re-created using the
multiplication of the default parameters for the maximal edge length provided
by the NETGEN-1D-2D-3D algorithm by factors: $2$, $4$ and $8$.}. The surface
mesh densities of the sphere correspond to the case labels \emph{coarse},
\emph{fine} and \emph{very fine} used in the Figures
\ref{fig:recError:sphere:NAG} and \ref{fig:recError:sphere:generalizedLsq}.
The \emph{fine} sphere mesh is chosen for further results comparison.

\begin{figure}
   {\footnotesize
   \begin{center}
   \begin{subfigure}[]{0.45\columnwidth}
       \centering
       \includegraphics[width=0.6\columnwidth]{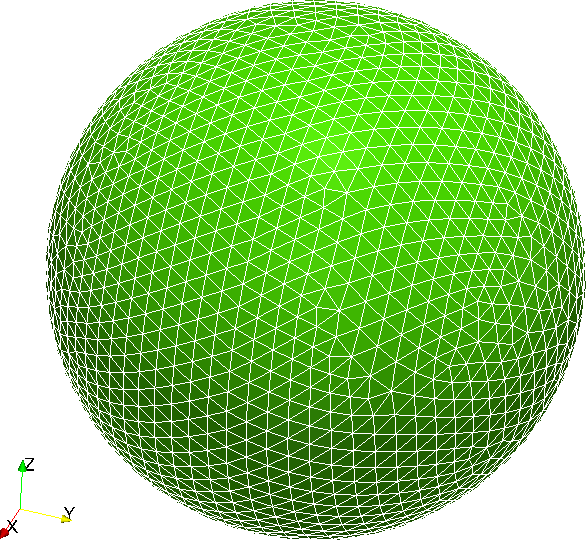}
       \caption{Sphere tool mesh (\emph{fine} case)}
       \label{fig:recResults:sphereMesh}
   \end{subfigure}
   \end{center}
   \begin{subfigure}[b]{0.45\columnwidth}
       \centering
       \includegraphics[width=0.6\columnwidth]{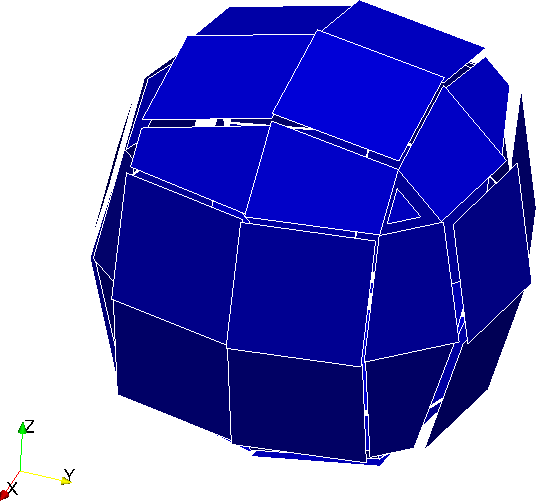}
   \end{subfigure}
   \begin{subfigure}[b]{0.45\columnwidth}
       \centering
       \includegraphics[width=0.6\columnwidth]{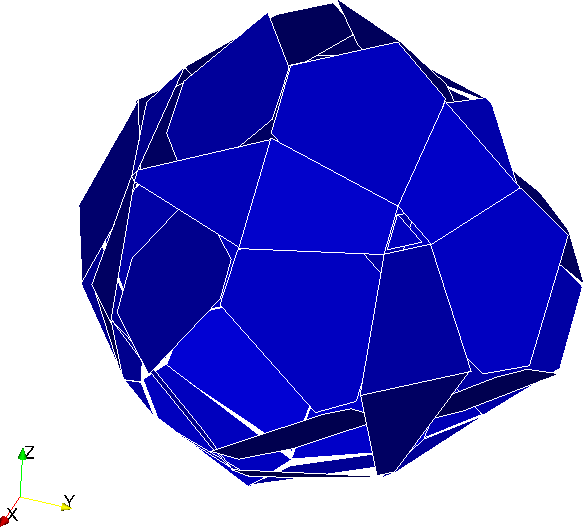}
   \end{subfigure}

   \begin{subfigure}[b]{0.45\columnwidth}
       \centering
       \includegraphics[width=0.6\columnwidth]{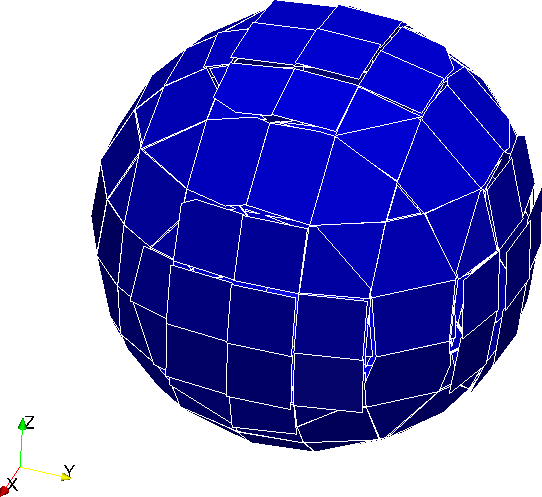}
   \end{subfigure}
   \begin{subfigure}[b]{0.45\columnwidth}
       \centering
       \includegraphics[width=0.6\columnwidth]{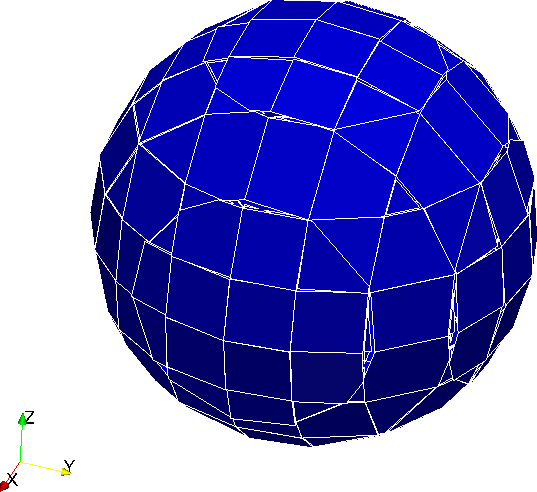}
   \end{subfigure}

   \begin{subfigure}[b]{0.45\columnwidth}
       \centering
       \includegraphics[width=0.6\columnwidth]{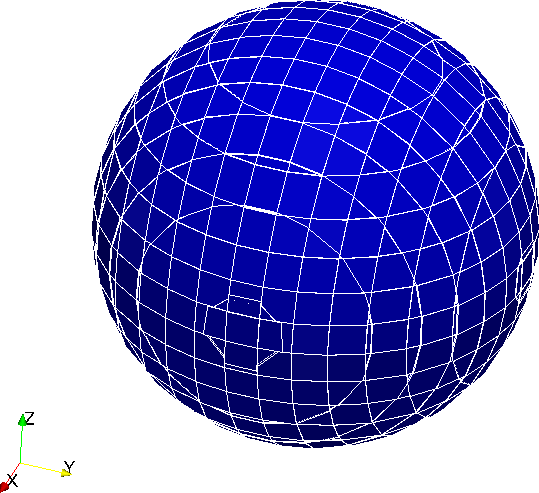}
   \end{subfigure}
   \begin{subfigure}[b]{0.45\columnwidth}
       \centering
       \includegraphics[width=0.6\columnwidth]{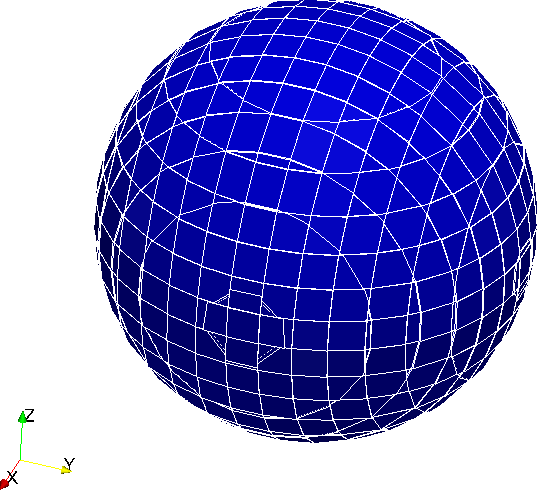}
   \end{subfigure}

   \begin{subfigure}[b]{0.45\columnwidth}
       \centering
       \includegraphics[width=0.6\columnwidth]{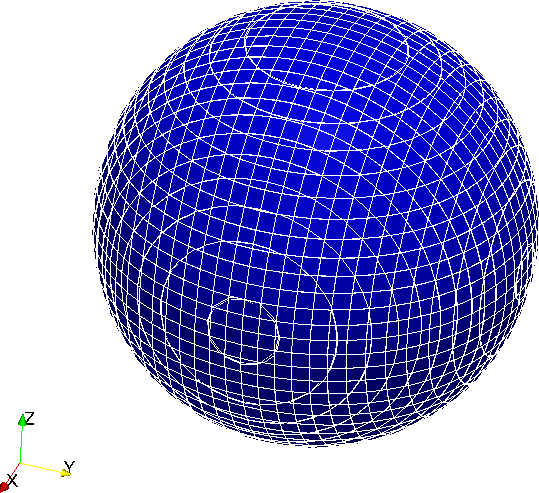}
       \caption{NAG gradient}
   \end{subfigure}
   \begin{subfigure}[b]{0.45\columnwidth}
       \centering
       \includegraphics[width=0.6\columnwidth]{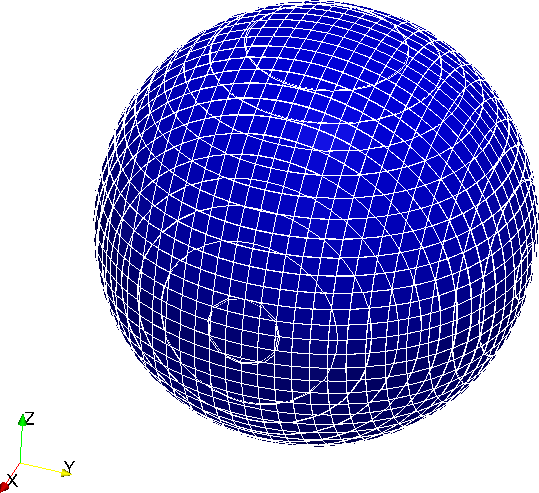}
       \caption{generalized LSQ gradient}
   \end{subfigure}

    \caption{Reconstructed spherical interface on an unstructured hexahedral
mesh. Interfaces are reconstructed on base meshes with densities: $4^3$, $8^3$,
$16^3$ and $32^3$} 

    \label{fig:recResults:comparison}
   }
\end{figure}

Figure \ref{fig:recResults:comparison} shows the input mesh for the sphere
(\emph{fine} case) and the comparison of the corresponding reconstructed
interfaces using both the NAG  gradient and the generalized LSQ gradient on
unstructured hexahedral meshes. The NAG gradient shows better results on the
coarsest hexahedral mesh in the volume of symmetric difference error.  This can
be seen when the absolute values are compared for the coarsest \emph{base
meshes} shown in Figures \ref{fig:recError:sphere:NAG} and
\ref{fig:recError:sphere:generalizedLsq}. Also, the visual appearance of the
interface geometry on the coarsest mesh is clearly better when the NAG gradient
is applied on an unstructured hexahedral mesh, as shown in Figure
\ref{fig:recResults:comparison}.

\begin{figure}
   {\footnotesize
   \begin{center}
   \begin{subfigure}[]{0.45\columnwidth}
       \centering
       \includegraphics[width=0.54\columnwidth]{figures/sphereFineComplete.png}
       \caption{Sphere tool mesh (\emph{fine} case)}
       \label{fig:recResults:sphereMeshTet}
   \end{subfigure}
   \end{center}

   \begin{subfigure}[b]{0.45\columnwidth}
       \centering
       \includegraphics[width=0.6\columnwidth]{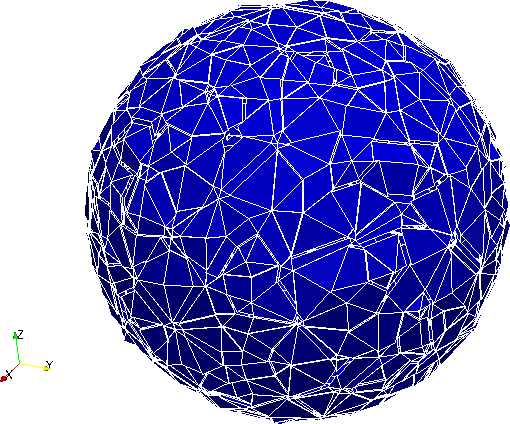}
       \caption{NAG gradient} 
   \end{subfigure}
   \begin{subfigure}[b]{0.45\columnwidth}
       \centering
       \includegraphics[width=0.6\columnwidth]{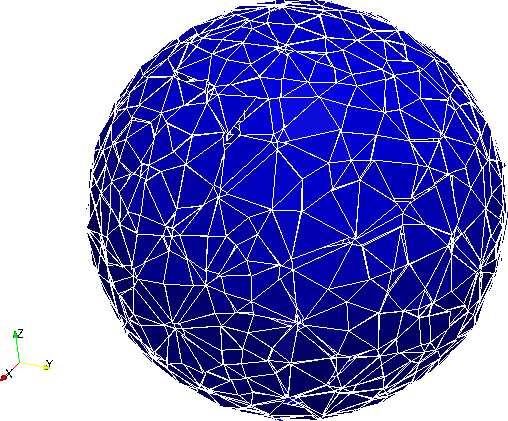}
       \caption{Generalized LSQ gradient}
   \end{subfigure}

    \caption{Reconstructed spherical interface on an unstructured tetrahedral 
mesh. Interfaces are reconstructed on a base mesh with $8774$ volumes}
    \label{fig:recResults:comparisonTet}
   }
\end{figure}

%
%
\begin{figure}
	\centering
		\includegraphics{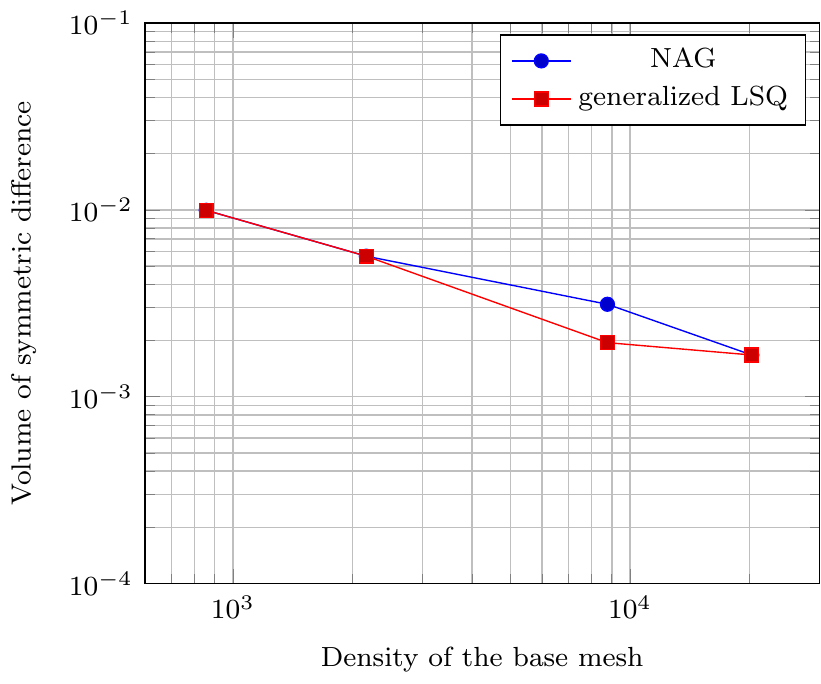}
    \caption{Reconstruction error for a sphere on a tetrahedral mesh with following numbers of volumes: 857, 2168, 8774, 20263}
    \label{fig:recError:sphere:tetra}
\end{figure}

Figure \ref{fig:recResults:comparisonTet} shows the reconstructed spherical
interfaces on a tetrahedral mesh. For this case the error magnitude is lower
when the NAG gradient is applied. Interface polygons on a tetrahedral mesh are
significantly disconnected, especially around the reconstruction points where
multiple tetrahedra meet in a mesh vertex. Around those points the volume
fraction values will be either very small ($\alpha_1 \approx 0$) or very large
($\alpha_1 \approx 1$) in which cases even the smallest perturbation in the
orientation of the interface normal causes significant disconnection of the
interface polygons. 

Figure \ref{fig:recError:sphere:tetra} shows the behavior of the interface
reconstruction error on an \emph{unstructured tetrahedral mesh}. The evaluation
of the reconstruction error is done for both the NAG gradient and the
generalized LSQ gradient. Both gradients show similar behavior of the error
magnitude with respect to the increase of the mesh density. On a tetrahedral
mesh the evaluation of the error convergence is not possible, because of its
irregular nature. Still, the diagram shown in Figure
\ref{fig:recError:sphere:tetra} shows that the magnitude of the reconstruction
error does not decrease as fast as on the unstructured hexahedral mesh. This
fact brings us to conclusion that the unstructured hexahedral mesh makes a
better choice for domain discretization for our geometrical VoF method. 

A polyhedral mesh provides direct connectivity via finite volume faces to all
surrounding cells which makes the gradient calculation more exact and
straightforward since it relies on the connectivity which is already present
and involves all available cell neighbors. However, the quality of the
polyhedral mesh will rely on the way it is created, which is usually done by
agglomerating tetrahedral cells. When tetrahedral cells are agglomerated to
create polyhedra, severely non-convex cells with significant non-planarity of
faces, appear in the mesh if the original tetrahedral mesh is not produced with
the Delaunay algorithm. Severely non-convex cells introduce numerical errors
and are unusable together with the intersection algorithms used for the
geometrical VoF method.

\begin{figure}
   {\footnotesize
   \begin{center}
   \begin{subfigure}[]{0.5\columnwidth}
       \centering
       \includegraphics[width=0.9\columnwidth]{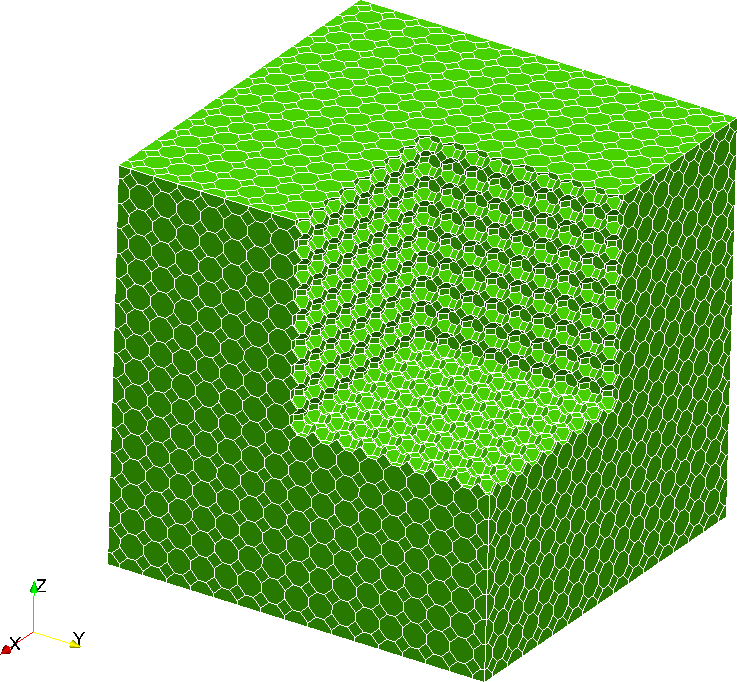}
       \caption{Mixed polyhedral/hexahedral base mesh}
       \label{fig:recResults:sphereMeshPoly}
   \end{subfigure}
   \end{center}

   \begin{subfigure}[b]{0.45\columnwidth}
       \centering
       \includegraphics[width=0.6\columnwidth]{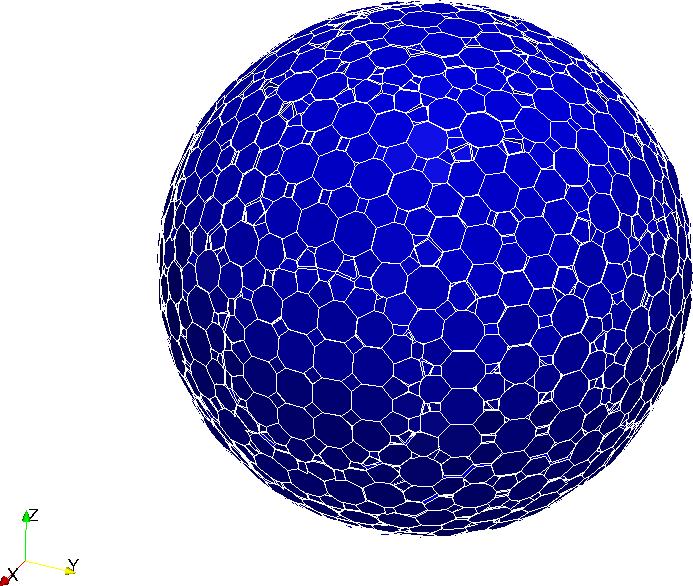}
       \caption{NAG gradient} 
   \end{subfigure}
   \begin{subfigure}[b]{0.45\columnwidth}
       \centering
       \includegraphics[width=0.6\columnwidth]{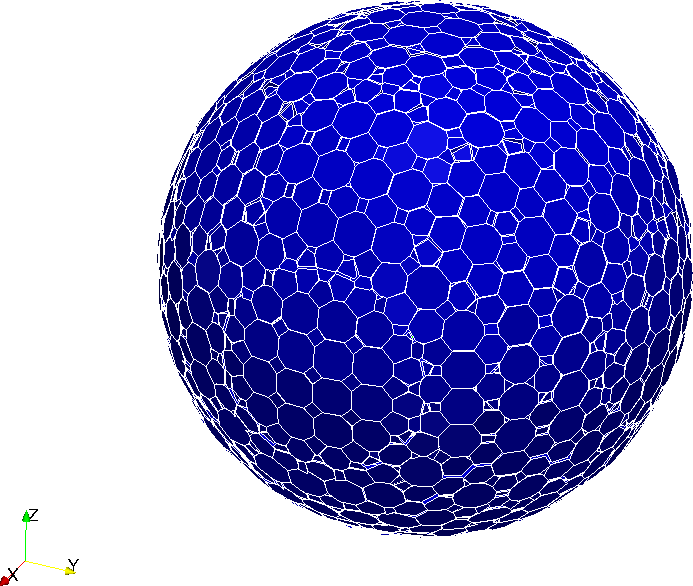}
       \caption{Generalized LSQ gradient}
   \end{subfigure}

    \caption{Reconstructed spherical interface on a mixed polyhedral mesh with $22065$ cells} 

    \label{fig:recResults:comparisonPoly}
   }
\end{figure}

To the best of our knowledge, a dynamic mesh refinement algorithm based
exclusively on polyhedral cells does not exist. Since we are interested in a
more accurate spatial resolution in the surrounding of a \emph{moving fluid
interface} (achieved by employing local dynamic AMR), the polyhedral mesh
results are shown only to prove the generality of the implemented geometrical
operations with respect to different cell shapes. 
\ref{fig:recResults:comparisonPoly} shows the reconstruction of a spherical
interface on a polyhedral base mesh which is created by agglomerating the
tetrahedra of a mesh discretized using equal sized hexahedral cells. The
hexahedral cells are then decomposed, each into $24$ tetrahedrons, which are
then agglomerated into polyhedra to ensure the convexity of the cells. This
mesh is then a mixture of hexahedral and polyhedral cells which is clearly
visible in the reconstructed interfaces. 

The behavior of the reconstruction error shows that there is a significant
accuracy gain in reconstructing the interface on an unstructured hexahedral
mesh, compared to the unstructured tetrahedral mesh. This type of mesh supports
local dynamic AMR within the frame of the existing topological mesh operations
in OpenFOAM: hexahedral refinement of cells. Moreover, the automatic mesh
generation of unstructured hexahedral meshes is widely available (e.g. via the
\emph{snappyHexMesh} tool of the OpenFOAM library). Hence, the unstructured
hexahedral mesh presents a natural choice for the domain discretization of a
geometrical VoF method supporting local dynamic AMR.

\subsection{Verification of the volume fraction advection}

In this section the results of the geometrical advection algorithm for standard
test cases are shown. Standard test cases are used to validate mass/volume
conservation, numerical boundedness of the volume fraction field i.e.
$\alpha_1 \in [0 + \epsilon,1 - \epsilon]$, where $\epsilon$ is the
reconstruction tolerance, geometrical stability of the interface as well as its
\emph{physically consistent motion}. A physically consistent motion of the
interface is a motion during which no \emph{artificial separation} is present
(e.g.  \emph{flotsam}/\emph{jetsam}), appearance of \emph{wisps} and
\emph{artificial deformation} of the interface are reduced to a minimum. 

\begin{description}
 \item Artificial separation of the interface involves large separated parts of the
fluid (flotsam and jetsam), as well as small changes in the volume fraction
field (wisps).Flotsam and jetsam are large parts of the interface that separate
during the advection artificially, as defined by \citet{NohWoodwardSLIC1976}.
Wisps are, usually very small, numerically artificial, interface polygons that
appear sporadically near the transported interface, as well as within the bulk.
In effect, this means that there is an additional layer of cells to the actual
interface cell layer, where the values of the volume fraction are either very
close to one ($\alpha \approx 1$) or to zero ($\alpha \approx 0$) for cells
that are not interface cells. Visualization of bulk wisps, as well as their
generation will depend on the reconstruction tolerance of the algorithm.
Reducing the reconstruction tolerance, will make the actual bulk wisp cell a
full cell, on account of introducing an error in mass conservation for this
cell. The presence of wisps in the bulk of the flow is due to the
non-divergence-free nature of the discrete flow field, which we counter in our
algorithm by introducing a narrow layer of cells where the advection takes
place. 

\item Artificial deformation of the interface is an interface motion which, although
resulting with a mass conservative volume fraction field bounded between $0$
and $1$, imposes an artificial change in the velocity with which the interface
is locally advected.  This error stems directly from the way the volumetric
phase flux is computed. 
\end{description}

The standard $E_g$ geometrical advection error is defined as follows
\citep{Aulisa2007}:
\begin{equation}
        E_g = \sum_c V_c (\alpha_c(t) - \alpha_c(t_0)),
        \label{eq:error:geom}
\end{equation}
where $c$ marks the cell of the volume $V_c$, $t$ marks the time value for
which the error is computed and $t_0$ marks the initial (reference) time value.
This error can only be applied for advection test cases for which the interface
configuration at time $t$ corresponds to the configuration at the initial time
$t_0$. For such cases, a periodical velocity field is used for advection.
Examples of such test cases are described by various authors and we have
used the following test cases to verify the geometrical advection algorithm:

\begin{compactitem}
   \item periodical translation along spatial diagonal,  
   \item rotation,  
   \item shear,  
   \item deformation. 
\end{compactitem} 




\subsubsection{Translation and rotation test}

Apart from a sharp interface representation within a single layer of cells, the
geometrical VoF method significantly reduces the artificial deformation of the
interface compared with algebraic VoF methods. Since there is a contribution to 
the volumetric phase flux coming from the point-cell neighbor cells (skew cell 
neighbors), the motion of the interface in this direction will not be artificially 
changed to a large extent. If the flux contributions stemming from the skew cell 
neighbors are completely neglected, as this is the case for algebraic advection 
schemes,  interface will be advected adhering to mass conservation, but the 
interface shape will be artificially changed to a large extent. 

The translation test case consists of a sphere of radius $0.15$, placed in the
center of a unit length box-domain with increasing uniform density, and it is
advected with Courant numbers: $0.1$, $0.5$ and $0.75$. The sphere is
translated along the spatial diagonal from the initial position to the two
orners of the box-domain and returned to initial position at the box center.
The overall distance traveled by the sphere in this test case is approximately 10
sphere diameters.  

\begin{figure}
   {\footnotesize
   \centering
   \begin{subfigure}[b]{0.49\columnwidth}
       \centering
       \includegraphics[width=\columnwidth]{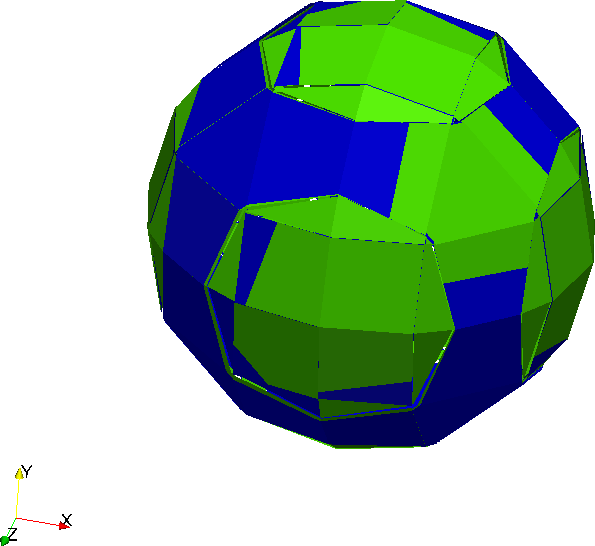}
       \caption{Mesh with $16^3$ volumes} 
   \end{subfigure}
   \begin{subfigure}[b]{0.49\columnwidth}
       \centering
       \includegraphics[width=\columnwidth]{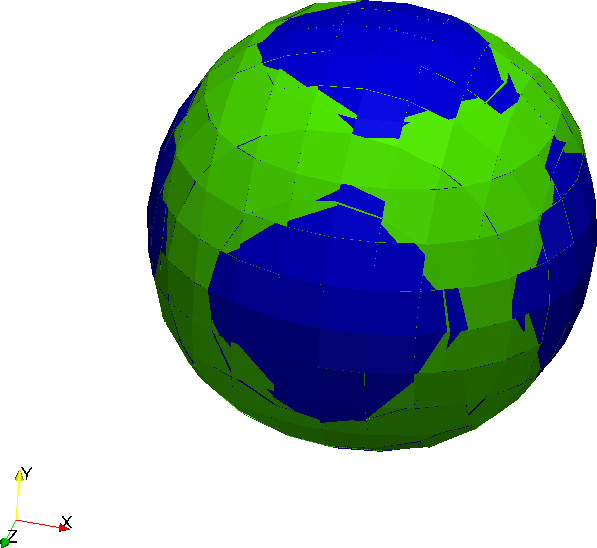}
       \caption{Mesh with $32^3$ volumes}
   \end{subfigure}
\vspace{0.5cm}

   \begin{subfigure}[b]{0.49\columnwidth}
       \centering
       \includegraphics[width=\columnwidth]{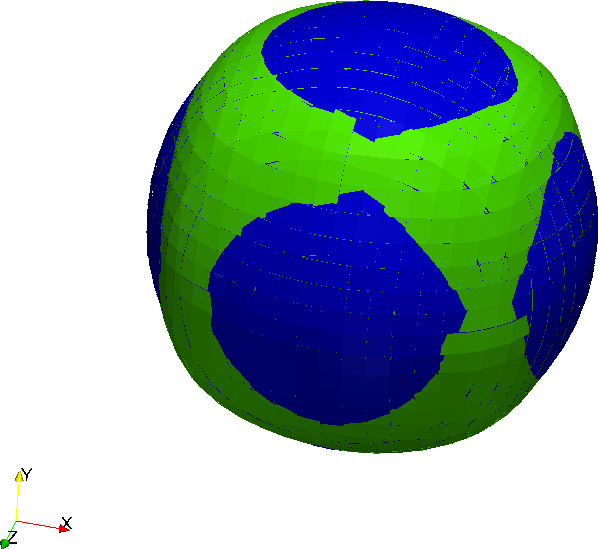}
       \caption{Mesh with $64^3$ volumes}
   \end{subfigure}
   \begin{subfigure}[b]{0.49\columnwidth}
       \centering
       \includegraphics[width=\columnwidth]{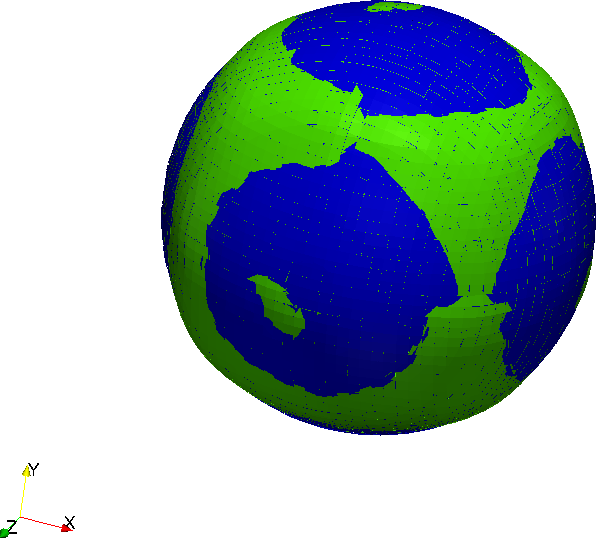}
       \caption{Mesh with $128^3$ volumes}
   \end{subfigure}

    \caption{Visual comparison of the initial (blue) and the final sphere shape
(green) for mesh different densities of the \emph{skew advection} test}

    \label{fig:skew:initialFinal}
   }
\end{figure}

Figure \ref{fig:skew:initialFinal} shows a comparison between the final
position of the sphere and the initial sphere position for increasing mesh
density and the Courant number of $0.5$. For an advection directed along the
spatial diagonal, the images show that the flux calculation, although fully
un-split, involves a slight deformation of the interface the skew direction.
Table \ref{tab:errors:translation} shows the geometrical error ($E_g$) for the
skew-translation test case. The relative mass conservation error values are
near machine tolerance for the constant translation velocity field and are thus
omitted. 

\begin{table} 
\footnotesize
\centering 
\caption{Advection errors of the periodical translation test case} 
\label{tab:errors:translation} 
\begin{tabular}{c l c c c c c} 
       \toprule
       Number of volumes && $16^3$ & $32^3$ & $64^3$ & $128^3$ & \\ 
        \hline
        \hline
      Courant number &&&&&& \\
      0.1  $$ && 5.32128e-04 & 4.24662e-04  & 8.76037e-04 & 6.99539e-04 & \\ 
      0.5  $$ && 2.75337e-04 & 3.85527e-04  & 7.54332e-04 & 4.39392e-04 &\\ 
      0.75 $$ && 3.15498e-04 & 4.10146e-04  & 6.86752e-04 & 5.82132e-04 &\\ 
        \bottomrule
\end{tabular}
\end{table}

\begin{table} 
\footnotesize
    \centering 
    \caption{Advection errors for the rotation test case} 
    \label{tab:errors:rotation} 
    \begin{tabular}{c  c c c c c} 
          \toprule
          Number of volumes & $16^3$ & $32^3$ & $64^3$ & $128^3$ & \\ 
          \hline
          \hline
          $E_g$ & 8.32758e-04 &  3.18881e-04 & 1.45851e-04 & 1.11573e-04 & \\ 
          \bottomrule
    \end{tabular}
\end{table}

\begin{figure}
   {\footnotesize
   \centering
   \begin{subfigure}[b]{0.49\columnwidth}
       \centering
       \includegraphics[width=\columnwidth]{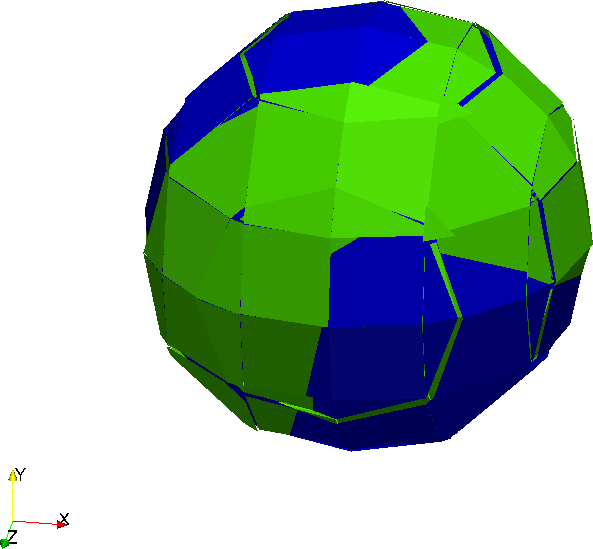}
       \caption{Mesh with $16^3$ volumes} 
   \end{subfigure}
   \begin{subfigure}[b]{0.49\columnwidth}
       \centering
       \includegraphics[width=\columnwidth]{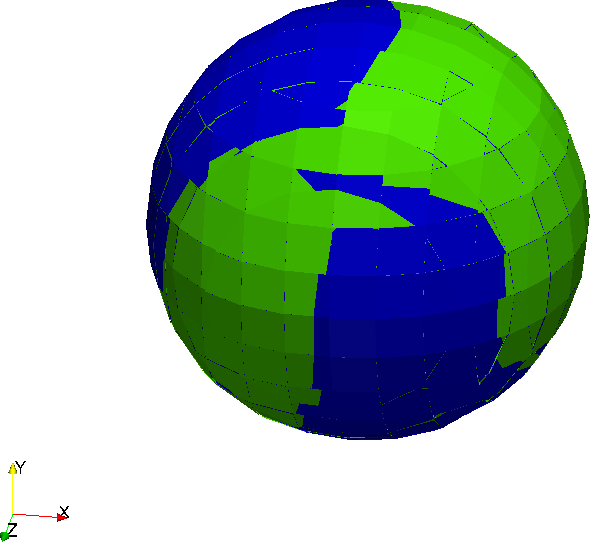}
       \caption{Mesh with $32^3$ volumes}
   \end{subfigure}

   \begin{subfigure}[b]{0.49\columnwidth}
       \centering
       \includegraphics[width=\columnwidth]{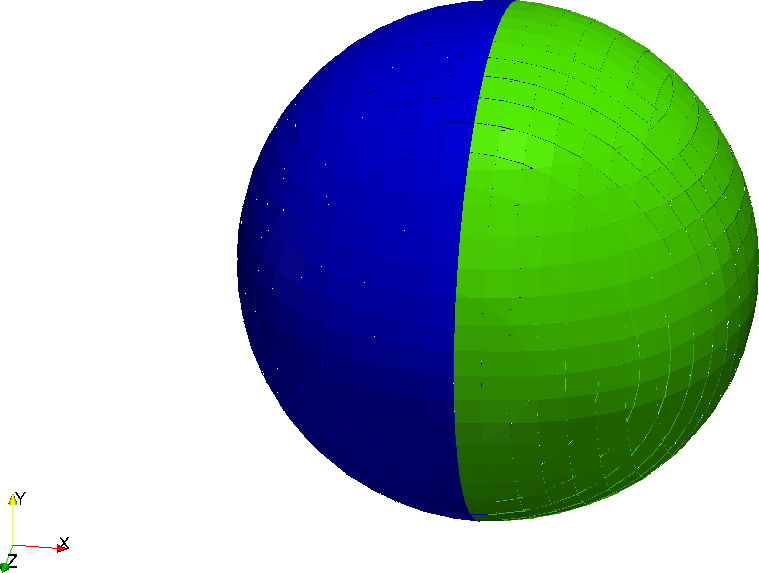}
       \caption{Mesh with $64^3$ volumes}
   \end{subfigure}
   \begin{subfigure}[b]{0.49\columnwidth}
       \centering
       \includegraphics[width=0.94\columnwidth]{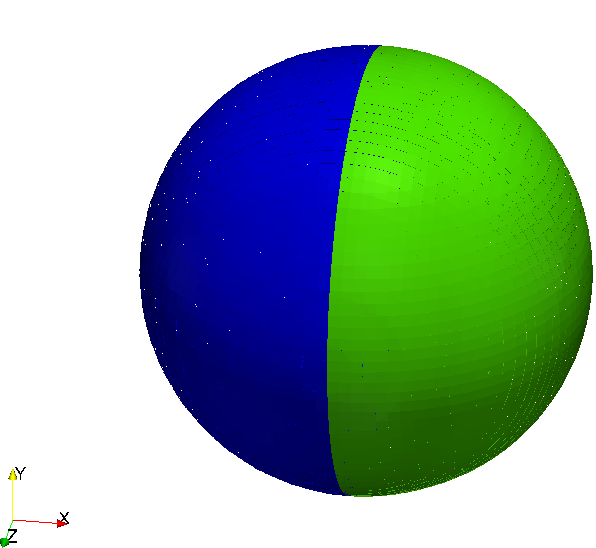}
       \caption{Mesh with $128^3$ volumes}
   \end{subfigure}

    \caption{Visual comparison of the intial (blue) and the final (green) sphere shape
 for different densities of the \emph{rotation} test}

    \label{fig:rotation:initialFinal}
   }
\end{figure}

The rotation test case consists of a sphere with the radius of $0.15$ placed at
$(0.5, 0.75, 0.5)$ within a unit-length box domain as described by
\citet{jofreUnstructuredPLIC}. The sphere is rotated with a Courant number
value of  $0.5$ based on the maximal velocity used to advect the interface. The
final and initial positions of the sphere are compared. Table
\ref{tab:errors:rotation} shows the behavior of the advection error $E_g$  
with respect to the increase of mesh density for the rotation test case. Figure
\ref{fig:rotation:initialFinal} shows the visual comparison between the final
and initial interface shape for different densities of the mesh. The rotation
test case involves computing intersections with very thin and distorted flux
polyhedra, and for this purpose the rotation test cases with mesh densities of
$64^3$ and $128^3$ were run in parallel such that the process boundary plane
lies orthogonal to the velocity field. This test stresses the robustness and
accuracy of the geometrical advection algorithm even for extreme shapes of very
thin flux polyhedra. 

The visual difference between the initial sphere on the left process boundary
(blue) and the final sphere on the right process boundary (green) shown on the
Figure \ref{fig:rotation:initialFinal} is not noticeable. For the coarser
meshes of $32^3$ and $16^3$ volumes, the overlap shown on a single process
domain is insignificant, which is confirmed by absolute values of the $E_g$
error. The error convergence behaves as reported by other authors
\citep{HernandezLopezPart1Advection2008}: the error converges faster on coarser
meshes and the magnitude of order of convergence magnitude is near $1$, since
it is defined by the convergence order of the Young's reconstruction algorithm.  

\subsubsection{Shear test case}

\begin{table} 
\footnotesize
\centering 
    \caption{Advection errors for the shear test case} 
    \label{tab:errors:shear} 
    \begin{tabular}{c c c c c c} 
          \toprule
          Number of volumes & $16^3$ & $32^3$ & $64^3$ & $128^3$ & \\ 
          \hline
          \hline
          Courant number &&&&& \\
          0.1   & 1.03638e-02 & 4.71128e-03 & 1.49640e-03 & 5.68667e-04 & \\ 
          0.5   & 1.11825e-02 & 5.78071e-03 & 2.59552e-03 & 1.18850e-03 & \\ 
          0.75  & 1.28960e-02 & 6.13972e-03 & 3.27130e-03 & 1.60405e-03 & \\ 
          \bottomrule
    \end{tabular}
\end{table}

\begin{figure}
   {\footnotesize
   \centering
   \begin{subfigure}[b]{0.49\textwidth}
       \centering
       \includegraphics[width=0.8\textwidth]{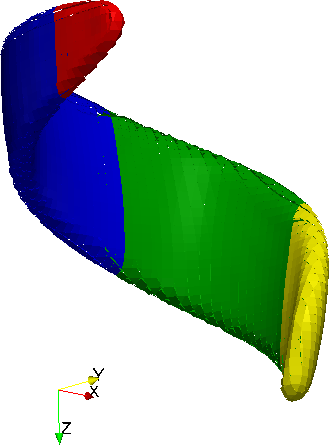}
       \caption{Extreme shear: colors mark parallel process domains} 
       \label{fig:shear:extreme}
   \end{subfigure}
   \begin{subfigure}[b]{0.49\textwidth}
       \centering
       \includegraphics[width=\textwidth]{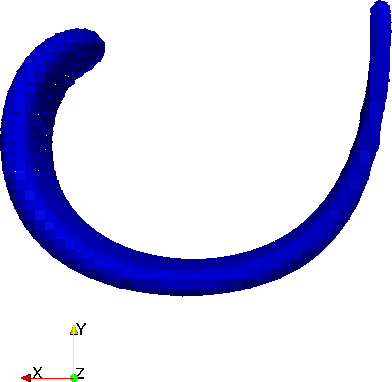}
       \label{fig:shear:top}
       \caption{Extreme shear: projection in the direction of the Z-axis}
   \end{subfigure}
\vspace{0.5cm}

   \begin{subfigure}[b]{0.5\textwidth}
       \centering
       \includegraphics[width=\textwidth]{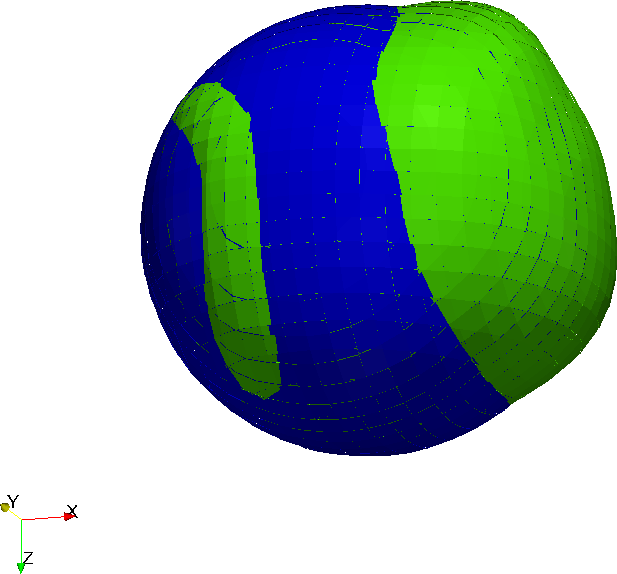}
       \caption{Initial (blue) and the final (green) sphere shape} 
       \label{fig:shear:comparison}
   \end{subfigure}

    \caption{Shear test case on a mesh with $64^3$ volumes}
    \label{fig:shear}
   }
\end{figure}

The shear test case configuration is described by \citet{LiovicCVTNA2006} in the following way:
\begin{align}
    \U(\x,\,t) = 
    \left(      
			\begin{array}{l}
				\sin(2\pi y) \sin(\pi x)^2 \cos \left(\frac{\pi t}{T}\right)\\
				-\sin(2\pi x) \sin(\pi y)^2 \cos \left(\frac{\pi t}{T}\right)\\
				u_\text{max} \left(1 - \frac{r}{R}\right)^2 \left(\frac{\pi t}{T}\right)
			\end{array}
    \right)
    \label{equation:shear-test-case}
\end{align}
We have used $T=3$, $u_\text{max} = 1$ and $R=0.5$ for the case settings to ensure
that the shear of the sphere will be extreme, which is visible in Figure
\ref{fig:shear:extreme}. The advection error $E_g$ of the shear test case is
shown on the Table \ref{tab:errors:shear}. The velocity field of the shear flow
test case is changing both in time and space, so the Courant number values are
computed for the maximal velocity in the flow domain for the current time step.
This Courant number value is then used to correct the time step for the next
iteration in order to maintain the prescribed fixed maximal Courant number. 

Figure \ref{fig:shear} shows the extreme state of the sphere in an imposed
shear flow, as well as the comparison between the initial and the final sphere
state for a mesh of $64^3$ volumes. The validation test cases with mesh
densities of $64^3$ and $128^3$ were run in parallel in order to validate the
computation of the volumetric phase flux across process domains for complex
flows. No differences in the results between the parallel and serial runs have 
been observed. The static parallel domain decomposition is shown in Figure
\ref{fig:shear:extreme} with different colours.

The quantitative results shown in the Table \ref{tab:errors:shear} as well as
the results shown in Figure \ref{fig:shear} present first (to the best of our
knowledge) numerically stable (mass conservative and numerically bounded)
results involving a three dimensional spatially complex and time-varying
velocity field used to advect the volume fraction with a directionally un-split
advection algorithm using a full discrete Lagrangian flow-map on an
unstructured mesh hexahedral mesh, in parallel.

\subsubsection{Deformation test case}

\begin{figure}
   {\footnotesize
   \centering
   \begin{subfigure}[b]{0.49\textwidth}
       \centering
       \includegraphics[width=\textwidth]{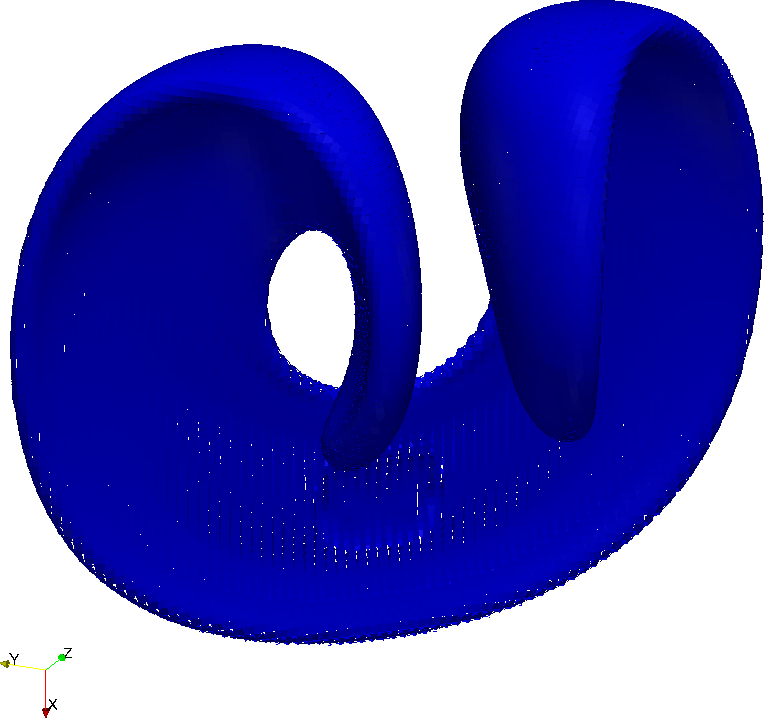}
       \caption{Extreme deformation of a sphere} 
       \label{fig:deformation:extreme}
   \end{subfigure}
   \begin{subfigure}[b]{0.5\textwidth}
       \centering
       \includegraphics[width=\textwidth]{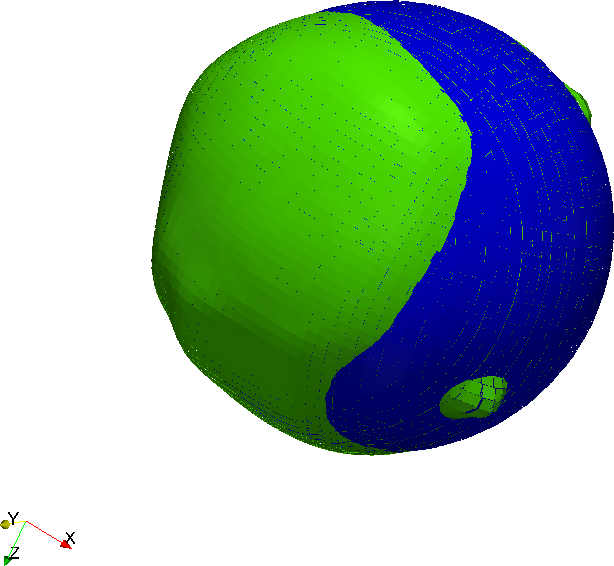}
       \caption{Initial (blue) and the final (green) sphere shape} 
       \label{fig:deformation:comparison}
   \end{subfigure}

       \caption{Deformation test case on a mesh with $128^3$ volumes}

    \label{fig:deformation}
   }
\end{figure}

The deformation test case, as described by
\citet{HernandezLopezPart1Advection2008}, reads 
\begin{align}
    \U(\x,\,t) = 
    \left( 
			\begin{array}{l}
			  2 \sin(2\pi y) \sin(\pi x)^2 \sin(2 \pi z) \cos \left(\frac{\pi t}{T}\right)\\
			  -\sin(2\pi x) \sin(\pi y)^2 \sin (2\pi z) \cos \left(\frac{\pi t}{T}\right)\\
			  -\sin(2\pi x) \sin(2 \pi y) \sin (\pi z)^2 \cos \left(\frac{\pi t}{T}\right)
			\end{array} 
    \right)
    \label{equation:deformation-test-case}
\end{align}
and we chose $T=3$ to ensure extreme deformation of the sphere. Table
\ref{tab:errors:deformation} shows the advection error values $E_g$ for
deformation test cases of different Courant numbers, $0.1$, $0.5$, and $0.75$,
as well as increasing mesh densities, $16^3$, $32^3$, $64^3$ and $128^3$.
Figure \ref{fig:deformation:comparison} shows a comparison between the initial
and final interface for the maximal Courant number $0.5$ and the mesh density
$128^3$, and Figure\ref{fig:deformation:extreme} shows the extreme deformation
of the sphere for the same parameter values. 

The geometrical instability present in the lower-middle part of the deformed
interface in the extreme shear state is visible in Figure
\ref{fig:deformation:extreme}. The cause of the instability lies in the fact
that the interface in this region becomes extended into a very narrow filament.
When the thickness of the layer approaches the cell size, the gradient
calculation of the Young's reconstruction method is not accurate enough to
stabilize the orientation of the interface normal. This is a common issue of
the geometrical VoF method and a detailed description is provided by
\citet{Cerne2002}. Two solutions may be applied in this case: a higher order
reconstruction and/or local dynamic AMR. A higher order reconstruction method
usually involves optimization of the interface normal orientation, which brings
additional computational cost to the overall algorithm. On the other hand, the
local dynamic AMR increases the accuracy of both the geometrical advection and
the flow solution. The \emph{locally increased mesh density } provided by local
dynamic AMR will enable the geometrical algorithm to resolve very thin
filaments, as well as the boundary layer surrounding the interface, while
keeping a \emph{low overall number of cells} in the mesh, thus increasing the
overall efficiency of the solution. 

\begin{table} 
\footnotesize
\centering 
\caption{Advection errors of deformation test case} 
\label{tab:errors:deformation} 
    \begin{tabular}{c c c c c c} 
          \toprule
          Number of volumes & $16^3$ & $32^3$ & $64^3$ & $128^3$ & \\ 
          \hline
          \hline
          Courant number &&&&& \\
          0.1   & 1.34430e-02 & 7.776114e-03 & 1.53454e-03 & 8.09107e-04 & \\  
          0.5   & 1.48808e-02 & 8.157570e-03 & 2.53035e-03 & 1.30304e-03 & \\  
          0.75  & 1.58007e-02 & 8.978900e-03 & 3.17220e-03 & 1.58943e-03 & \\ 
          \bottomrule
    \end{tabular}
\end{table}

\subsection{Results using local adaptive AMR}

\begin{figure}
   {\footnotesize
   \centering
   \begin{subfigure}[b]{0.43\textwidth}
       \centering
       \includegraphics[width=\textwidth]{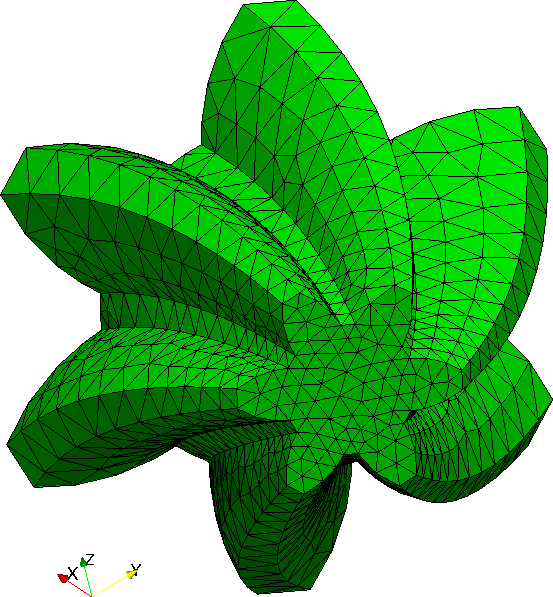}
       \caption{Input mesh (exact interface)} 
       \label{fig:helical:input}
   \end{subfigure}
   \quad
   \begin{subfigure}[b]{0.43\textwidth}
       \centering
       \includegraphics[width=\textwidth]{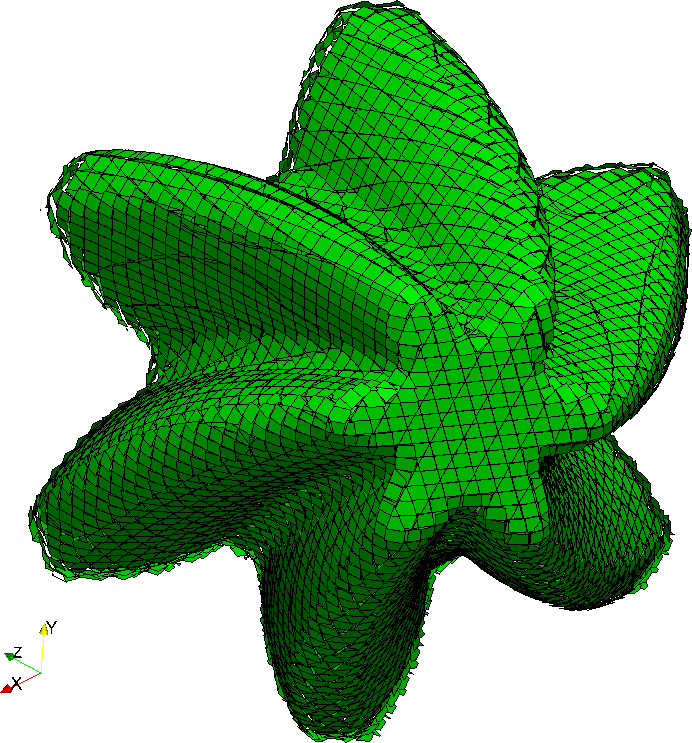}
       \caption{Reconstructed interface: front view}
       \label{fig:helical:final-front}
   \end{subfigure}
\vspace{0.5cm}

   \begin{subfigure}[b]{0.43\textwidth}
       \centering
       \includegraphics[width=\textwidth]{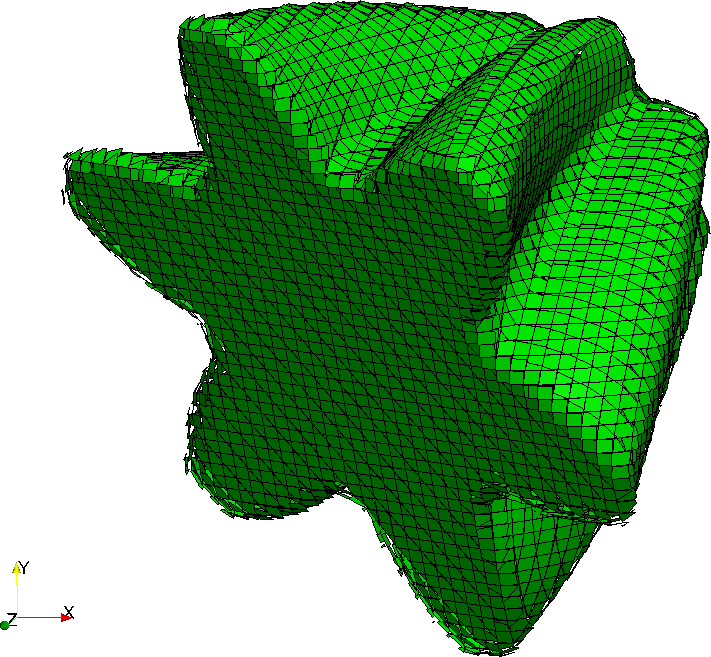}
       \caption{Reconstructed interface: back view}
       \label{fig:helical:final-back}
   \end{subfigure}
   \quad
   \begin{subfigure}[b]{0.43\textwidth}
       \centering
       \includegraphics[width=\textwidth]{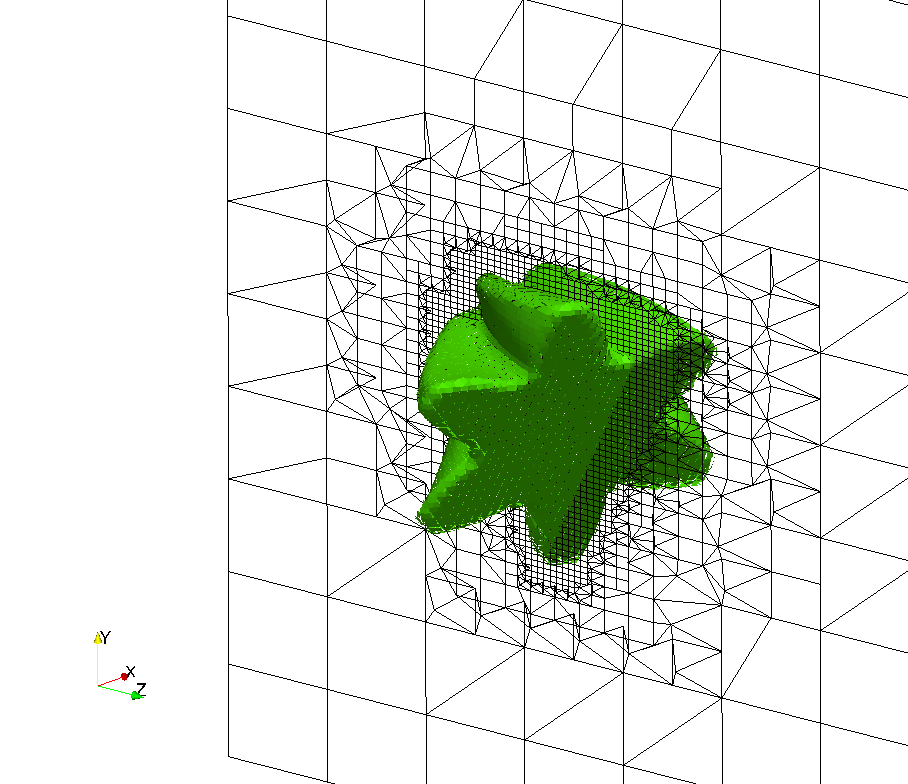}
       \caption{Background AMR refined mesh (wireframe) and reconstructed interface} 
       \label{fig:helical:final-background}
   \end{subfigure}

       \caption{Illustration of capability to reconstruct complex interface geometries: Interface 
         reconstruction for a helical gear geometry using local dynamic AMR}

    \label{fig:helical}
   }
\end{figure}

\subsubsection{Reconstruction}

To emphasize the advantages of basing our developments on an arbitrary
unstructured mesh, we have chosen a reconstruction verification case involving
a complex geometry. Arbitrarily, a geometry has been chosen consisting of a head 
part of a helical transmission gear for which the mesh was automatically generated. 
Clearly, this choice of the interface geometry is artificial, however it shows that 
arbitrary complexity of the interface reconstruction can be fully resolved with very 
little computational overhead using local dynamic AMR on unstructured hexahedral meshes, 
even with a geometrically first order convergent reconstruction algorithm.

Figure \ref{fig:helical:input} shows the prescribed interface in the form of a
helical gear (input mesh). The interface reconstructed with a refinement
level of $4$ is shown in the front view in Figure \ref{fig:helical:final-front}
and from the back view in Figure \ref{fig:helical:final-back}. The background mesh
of together with the refinement layers are visible in Figure
\ref{fig:helical:final-background}. Note that the initial mesh is very coarse;
the length scale of a cell is in the order of magnitude of the width of the
gear. 

Table \ref{tab:amr:reconstruction} contains the volume of symmetric difference
error $E_{vsd}$, order of error convergence, as well as normalized CPU time
and the normalized number of interface polygons with increasing level of mesh
refinement. The small order of convergence observed initially for the refinement 
level $1$ comes from the fact that the interface is severly under-resolved on the
initial very coarse  mesh. Once the local refinement reaches the cell length
scale that is enough to sufficiently resolve the curvature of the interface,
the convergence of the error increases significantly, which is expected for the
geometrically first order convergent Youngs reconstruction algorithm. With
further local mesh refinement the order of convergence of the $E_{vsd}$ error
stabilizes. 

\begin{figure}
	\centering
		\includegraphics{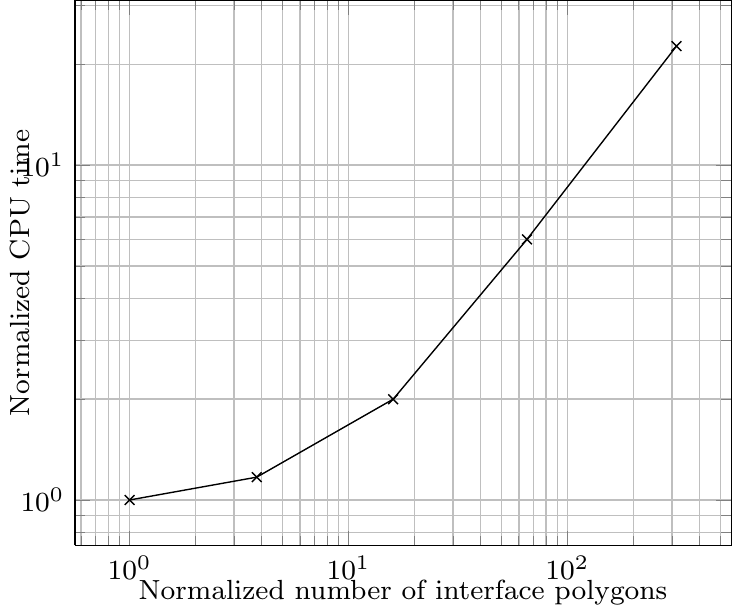}
     \caption{Linear dependency of the CPU time with respect to the number of interface polygons}
     \label{fig:reconstruction-cputime}
\end{figure}

The CPU time increases almost linearly with the increase in the number of
interface polygons, thus defining the overall reconstruction algorithm linear
in complexity $O(N_{polygons})$. This linear relationship is visualized in
Figure \ref{fig:reconstruction-cputime} in the form of a diagram.

\begin{table} 
\footnotesize
\centering 
\caption{Reconstruction data with local dynamic AMR} 
\label{tab:amr:reconstruction} 
    \begin{tabular}{l c c c c c } 
          \toprule
          Refinement level & 0 & 1 & 2 & 3 & 4\\ 
          \hline
          \hline
          $E_{vsd}$                             & 5.36139e-06 & 3.79479e-06 & 1.37713e-06 & 3.85176e-07 & 1.27413e-07 \\  
          $E_{vsd}$ order of convergence        & n/a & 0.49 & 1.46 & 1.83  & 1.59 \\  
          Normalized number of polygons         & 1 & 3.81 & 16 & 65.35  & 315 \\ 
          Normalized CPU time                   & 1 & 1.17 & 2 & 6 & 22.67 \\ 
          \bottomrule
    \end{tabular}
\end{table}

\subsubsection{Advection}

\begin{figure}
    \footnotesize
    \centering
     \includegraphics[width=0.5\textwidth]{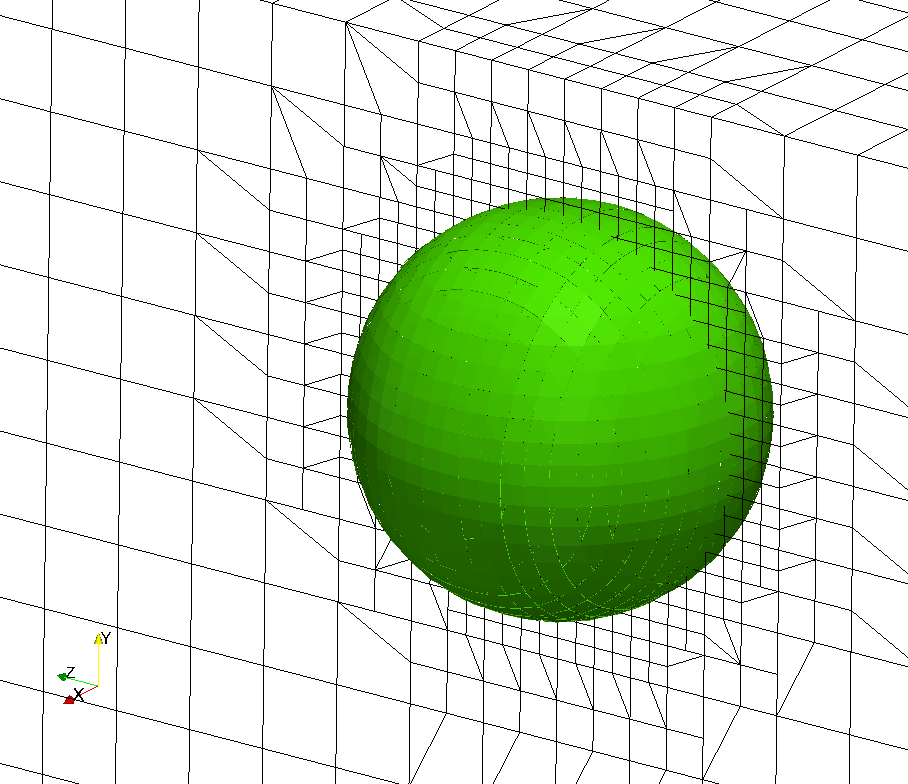}
     \caption{Final sphere shape for a rotation case using local dynamic AMR with level 2 refinement}
     \label{fig:amr:rotation}
\end{figure}

In \citep*{MaricVoFoamMapping2013} we have presented the
results for the shear test case together with local dynamic AMR to verify the
geometrical mapping algorithm. Table \ref{tab:amr:shear} holds an excerpt of
the solution results for this test case. The results of the shear advection
executed together with the local dynamic AMR on an initial mesh of $16^3$
volumes are consistent with the results obtained for the corresponding
uniformly refined meshes with densities defined by the finest refinement level.
The normalized CPU time with respect to the corresponding uniform shear test
case shows the benefit of local dynamic AMR when it comes to the computational
cost. To obtain the magnitude of the volume of symmetric error $E_{vsd}$
corresponding to the mesh of $128^3$ volumes (refinement level 3), only 58\% of
the CPU time is required. Further speedup is to be achieved by refactoring and
optimization of the geometrical library. Figure \ref{fig:amr:rotation} shows the
final result of the rotation test case with local dynamic AMR (level 2
refinement).

\begin{table} 
\footnotesize
\centering 
\caption{Shear case data with local dynamic AMR} 
\label{tab:amr:shear} 
    \begin{tabular}{l c c c} 
          \toprule
          Refinement level & 1 & 2 & 3 \\ 
          \hline
          \hline
          $E_{vsd}$                             & 5.8131E-003 & 2.7268E-003 & 1.11965E-003 \\  
          $E_{vsd}$ order of convergence        & 0.94 & 1.09 & 1.28 \\  
          Normalized CPU time                   & 0.83  & 0.71 &  0.58 \\ 
          \bottomrule
    \end{tabular}
\end{table}

\section{Summary \& Conclusions}
%
%
We provide considerable detail on the characteristics and fundamental steps of method 
development of a novel geometrical unsplit VoF algorithm for arbitrary unstructured 
meshes. Our algorithm has shown to produce a bounded volume fraction field; moreover, 
it is mass conservative and robust, even though its advection part is based on 
a discrete Lagrangian flow map (velocities in mesh points), which has been 
reported to be unfeasible in literature so far. 

%
%
The method's advection algorithm -- besides its use of an unique discrete 
Lagrangian flow map -- essentially comprises of a novel flux correction 
algorithm and narrow band computation: a fast iterative flux correction 
algorithm was implemented as well as an efficient narrow band computation 
based exclusively on the volume fraction field. This has been accomplished 
without introducing additional data structures and related algorithms 
required to update the narrow band as the interface moves. %
The key feature of the reconstruction algorithm is its accurate Node Averaged Gauss 
method used for calculating the phase fraction gradient, which results in lowest 
reconstruction errors reported so far for unstructured meshes.

The implementation of the geometrical intersection algorithms used both for 
reconstruction and advection is completely general, which makes the overall 
framework applicable to arbitrary cell types. %
However, the results show the algorithm to be more accurate on
unstructured hexahedral meshes than on tetrahedral meshes. In order to obtain
more accurate solutions on a tetrahedral mesh, higher order reconstruction
and/or advection algorithms will be required, which is to be expected. 
To increase the overall accuracy on hexahedral meshes, we have
extended the local dynamic Adaptive Mesh Refinement (AMR) engine in OpenFOAM
and implemented a geometrical mapping algorithm for the volume fraction field.
Results obtained using the adaptive geometrical VoF algorithm are promising and
are showing expected behaviour in absolute error values as well as error
convergence with a low computational overhead when compared to static mesh 
cases.  

The implementation of the geometrical library and the client simulation 
applications was done in a modular way using a policy based design in the C++ 
programming language, allowing for straightforward future extensions, if proven 
to be necessary. %
Our VoF algorithm is fully parallelized using the domain decomposition approach 
and is developed such that it is fully dimension agnostic following the design 
of the OpenFOAM library, which means that the transport automatically supports 
both two and three dimensional computation. The results are based on the majority 
of the standard established test cases for verification and validation of VoF 
algorithms and show that the new geometrical VoF algorithm delivers accurate and 
reliable results on unstructured meshes, even for large Courant numbers and 
complex velocity fields.

Our further research is directed towards the coupling of the geometrical VoF
algorithm with the pseudo-staggered flow solution algorithms in OpenFOAM on
arbitrary unstructured meshes. Another point of interest related to a
geometrically transported volume fraction field is the balancing of 
interfacial forces which will enable simulations of surface-tension
dominated flows in flow domains of arbitrary geometrical complexity.

\bibliographystyle{elsarticle-harv}
\bibliography{bibliography}

\end{document}